# On Hyperelastic Crease

Siyuan Song, Mrityunjay Kothari*, Kyung-Suk Kim[†]

School of Engineering, Brown University, Providence, RI 02912, US
[†]Corresponding Author: kyung-suk_kim@brown.edu

Abstract

We present analyses of crease-formation processes and stability criteria for general incompressible hyperelastic solids by solving generic singular perturbation problems. A generic singular perturbation over a laterally compressed hyperelastic half-space creates a far-field eigenmode of three energy-release angular sectors separated by two energy-elevating sectors of incremental deformation. The far-field eigenmode is found to brace the near-tip energy-release field of the surface flaw against the transition to a self-similar expansion field of creasing, and the braced-incremental-deformation (*bid*) field has a unique shape factor that determines the creasing transition stability. The shape factor is identified by two tangential-manifold conservation integrals that represent the geometry of a subsurface dislocation in the tangential manifold. The shape factor is a monotonically increasing function of compressive strain. Incompressible Mooney-Rivlin and neo-Hookean solids have the same shape-factor function. When the shape factor is below unity, the *bid* field is configurationally stable. However, the *bid* field undergoes a higher-order transition to a crease field at the crease limit point at which the shape factor becomes unity with a compressive strain of 0.356. At the crease-limit point, we have two asymptotic solutions of the crease-tip folding field and the leading-order far field with two separate scaling parameters. We could get the ratio between the two parameters with matched asymptotes. Our analyses show that the traction-free flat surface is stable against singular perturbation up to the crease limit point and becomes unstable beyond the crease limit point. However, the flat state is metastable against a regular perturbation between the crease limit point and Biot's 'wrinkle critical point' of surface-flatness collapse, which is a first-order instability point. We introduced a novel finite element method (FEM) for simulating the *bid* field in a hyperelastic half-space with a finite-size simulation domain. Furthermore, we uncovered with the Gent model (1996) that the strain-stiffening in hyperelasticity alters the dependence of the shape factor on the compressive strain, raising crease resistance. The new findings in hyperelastic crease mechanisms will help study ruga mechanics of self-organization and design soft-material structures and skin conditions for high crease resistance.

*Current address: Department of Mechanical Engineering, University of New Hampshire, Durham, NH, US

**Introduction**

Here, we consider the free-surface creasing process of a hyperelastic solid by solving a generic singular-field perturbation problem of nonlinear finite affine deformation in a hyperelastic half-space under lateral compression. Our formulation is for general incompressible hyperelastic solids. Nevertheless, we specifically include detailed formulations for incompressible neo-Hookean solids (Treloar, 1943; Rivlin, 1948) to present a transparent understanding of crease processes that can be compared with previous studies. Regarding the previous studies, extensive experimental testing (e.g., Cho & Gent, 1999; Trujillo et al., 2008; Jin et al., 2021) and computational analyses (e.g., Hohlfeld & Mahadevan, 2011; Hong et al., 2009; Yang et al., 2021) of hyperelastic free-surface creasing have been reported. Nonetheless, understanding the crease deformation-field characteristics and the configurational energetics of crease process has been still partial or incomplete. Studies on the deformation-field characterization include the near-tip asymptotic deformation- and stress-field analyses with crease-tip self-contact folding kinematics (Silling, 1991; Ciarletta, 2018 & 2020). Investigations on creasing processes comprise analyses of the crease limit point (Cao & Hutchinson, 2012; Diab et al., 2013; Diab & Kim, 2014; Ciarletta & Truskinovsky, 2019) and the critical point of surface-flatness collapse (Biot, 1963). The Biot's critical collapse point was found to be the unique first-order bifurcation point in an incompressible neo-Hookean half-space under lateral compression that represents the first-order transition of the flat-surface deformation into a crease configuration (Pandurangi et al., 2020 & 2022).

In this paper, regarding the singular perturbation, we consider various surface flaw types that induce intense near-tip incremental-deformation (Biot, 1965; Ogden, 1997) fields over the affine compressive deformation of the hyperelastic half-space under lateral compression. Our analysis reveals that the flaw-tip deformation field is braced by the far-field compression not to be transited to a crease deformation field until the far-field loading reaches the crease-limit point. Here, we name the flaw-tip incremental-deformation field prior to the crease limit point as a '*bid* field,' meaning 'braced incremental-deformation field.' We also named the transition point of the *bid* field to the crease field as the 'crease limit point,' delineating it from the Biot's 'wrinkle critical point' of surface-flatness collapse.

In the following sections, our singular perturbation analysis reveals how the far-field incremental deformation braces the flaw-tip field against transition. To this end, we introduce a notion of nominally linear-elastic incremental-deformation response and utilize duel-frame harmonic characteristics of the incremental field to identify the eigenmode of the far field that braces the near-tip field. Then, relevant novel path-independent integrals applicable to a tangential-manifold (TM) field of an affine nonlinear finite deformation are introduced to characterize the *bid* field and its transition behavior. In this process, we found that the *bid* field is represented by an analytically-continued TM subsurface-dislocation field denoted as a $S^{(bid)}$ field, where $S^{(bid)}$ is a well-defined *bid*-field shape factor which is expressed by the TM-dislocation quantities. The shape factor, $S^{(bid)}$, determines the stability of the higher-order transition at the crease limit point. Along with the theoretical advancement, we also developed a novel finite element method (FEM) for simulating the *bid* field in a hyperelastic half-space with a finite simulation domain. All these developments allowed us to complete the matched asymptotes of a crease field in incompressible neo-Hookean half-space and investigate the effects of strain stiffening on crease resistance with a Gent model (1996), which was first reported by Chen et al. (2014).

**Mathematical Formulation**

In this section, we present the mathematical structure of the far-field incremental deformation caused by surface flaws, including a crease, solving a singular perturbation problem in highly nonlinear hyperelastic half-space under lateral compression. Figure 1(a) illustrates a schematic of near, intermediate, and far fields of crease deformation to match interior and exterior perturbation asymptotes in a hyperelastic half-space. Here, we treat the crease field as a subclass of *bid* fields of surface flaws caused by lateral compression. As a result, we articulate robust computational formulas that characterize hyperelastic creasing processes. To this end, we consider incremental deformation of a general incompressible hyperelastic solid, decomposing the deformation in three steps, as shown in Fig.1(b). In the first step, a stress-free half-space, $X_2 \leq 0$, body (A) is uniformly compressed to (B) with a deformation gradient $\mathbf{F}^{[0]}$ of a lateral stretch $\lambda$ by a uniform stress $\boldsymbol{\sigma}^{(B)}$. In this step, a material point $\mathbf{X}$ in the stress-free reference configuration (A) is mapped to $\boldsymbol{x}$ in (B). In the second step, the configuration (B) is perturbed by $\mathbf{F}^{[1]}$ to (C) with displacement $\boldsymbol{u}(\boldsymbol{x})(=\boldsymbol{x}^\dagger - \boldsymbol{x})$, where $\boldsymbol{x}$ and $\boldsymbol{x}^\dagger$ respectively represent the material points in states (B) and (C). In the third step, the Cauchy stress $\boldsymbol{\sigma}$ in (C) is transformed into the nominal (or Piola) stress $\widetilde{\boldsymbol{\sigma}}$ in (B). Here, we define the nominal displacement gradient field $\nabla_x \widehat{\boldsymbol{u}}(\boldsymbol{x})$ of an incrementally linear-elastic tangential manifold (TM) by connecting $\widetilde{\boldsymbol{\sigma}}(\boldsymbol{x})$ to $\nabla_x \widehat{\boldsymbol{u}}(\boldsymbol{x})$ with the tangential compliance at (B) as depicted in (E), where $\widehat{\boldsymbol{u}}(\boldsymbol{x}) = \boldsymbol{x}^\dagger - \boldsymbol{X}^{[\mathrm{TM}]}$ for a TM stress-free material point $\boldsymbol{X}^{[\mathrm{TM}]}$ at (D). Then, we use various asymptotically valid path integrals in the TM field to analyze the *bid*-field characteristics. Hereafter, we use Italic and Greek subscripts for 3D and 2D indices with summation convention, respectively, and $'$ for differentiation.

**M.1 Incremental-elasticity characteristics of hyperelastic deformation**

The incremental displacement $\boldsymbol{u}$ and the incremental deformation gradient $\delta \mathbf{F}$ are interrelated as

$$\mathbf{F} = \mathbf{F}^{[1]}\mathbf{F}^{[0]} = \mathbf{F}^{[0]} + \delta \mathbf{F} \tag{1}$$

for which $\mathbf{F}^{[1]} = \mathbf{I} + \nabla_x \boldsymbol{u}$, with identity tensor $\mathbf{I}$, that leads to

$$\delta \mathbf{F} = \nabla_x \boldsymbol{u}\, \mathbf{F}^{[0]}. \tag{2}$$

As $\delta \mathbf{F}$ is work conjugate of the nominal stress $\mathbf{P}$ in the reference configuration at (A), we have

$$\delta w = tr(\mathbf{P}\delta \mathbf{F}^T) = tr\left[\mathbf{P}\, \mathbf{F}^{[0]^T}(\nabla_x \boldsymbol{u})^\mathrm{T}\right] \tag{3}$$

where $w(\mathbf{F})$ is the strain energy per volume at (A) in Fig. 1(b), and (3) is further reduced to

$$\delta w = tr\left[\mathrm{J}^{[0]}\mathrm{J}^{[1]}\boldsymbol{\sigma}\, \mathbf{F}^{[1]^{-T}}(\nabla_x \boldsymbol{u})^\mathrm{T}\right] = tr\left[\mathrm{J}^{[0]}\mathbf{P}^{[1]}(\nabla_x \boldsymbol{u})^\mathrm{T}\right] = tr[\widetilde{\boldsymbol{\sigma}}(\nabla_x \boldsymbol{u})^\mathrm{T}] \tag{4}$$

for $J = J^{[0]}J^{[1]}$, Jacobian of the mapping $\mathbf{X} \to \mathbf{x}^\dagger$. We call the work conjugate of $\nabla_x \mathbf{u}$, $\widetilde{\boldsymbol{\sigma}} = J^{[0]}\mathbf{P}^{[1]}$, 'unperturbed-configuration (UPC) nominal stress' at (B). Here, $\boldsymbol{\sigma}$ is Cauchy stress at (C).

From now on, we denote the tangential stiffness of the hyperelastic body at (B) for the strain energy per reference volume at (A), $\widetilde{w}(\nabla_x \mathbf{u})$ for which $\delta \widetilde{w}(\nabla_x \mathbf{u}) = \delta w((\mathbf{I} + \nabla_x \mathbf{u})\mathbf{F}^{[0]})$, as

$$\mathcal{A}_{ijkl} = \left. \frac{\partial^2 \widetilde{w}}{\partial u_{i,j} \, \partial u_{k,l}} \right|_{(B)}. \tag{5a}$$

for hyperelastic solids. The nominal stiffness tensor $\mathcal{A}_{ijkl}$ is not symmetric for the indices $(i,j)$ and $(k,l)$ respectively, but has major symmetry for the $\{(i,j),(k,l)\}$ index-group pair. For incompressible solids, $\delta \widetilde{\boldsymbol{\sigma}} = \mathcal{A} : \nabla_x \mathbf{u} - \delta p \mathbf{I}$ for which $p$ denotes pressure. Similarly, we introduce the tangential compliance of the hyperelastic body at (B) as

$$\mathcal{S}_{ijkl} = \left. \frac{\partial^2 \widetilde{w}^{(c)}}{\partial \widetilde{\sigma}_{ij} \, \partial \widetilde{\sigma}_{kl}} \right|_{(B)} \tag{5b}$$

where $\widetilde{w}^{(c)}(\widetilde{\boldsymbol{\sigma}})$ is the complementary strain energy of the hyperelastic body. Then, $\mathcal{S} = \mathcal{A}^{-1}$, and if the body is incompressible, $\nabla_x \mathbf{u} = \mathcal{S} : (\delta \widetilde{\boldsymbol{\sigma}} + \delta p \mathbf{I}) = \mathcal{S} : \delta \widetilde{\boldsymbol{\sigma}}$.

Hereafter, we represent TM quantities with '^'. Thenceforth, the TM strain-energy $\widehat{w}$ and the TM complementary-strain-energy $\widehat{w}^{(c)}$ per stress-free configurational volume at (D) for the incrementally linear-elastic TM displacement gradient $\nabla_x \widehat{\mathbf{u}}(x)$ $(= \nabla_x \mathbf{u} + \mathcal{S} : \boldsymbol{\sigma}^{(B)}$ for $|\nabla_x \mathbf{u}| \ll 1$ ) and the TM stress $\widehat{\boldsymbol{\sigma}}$ of $\{\nabla_x \widehat{\mathbf{u}} = \mathcal{S} : \widehat{\boldsymbol{\sigma}}\}$ become

$$\widehat{w}(\nabla_x \widehat{\mathbf{u}}) = \frac{1}{2}(\nabla_x \widehat{\mathbf{u}}) : \mathcal{A} : (\nabla_x \widehat{\mathbf{u}}) = \widehat{w}^{(c)}(\widehat{\boldsymbol{\sigma}}) = \frac{1}{2}\widehat{\boldsymbol{\sigma}} : \mathcal{S} : \widehat{\boldsymbol{\sigma}}. \tag{6a}$$

Here,

$$\delta \widehat{w}(\nabla_x \widehat{\mathbf{u}}) = \delta \widetilde{w}(\nabla_x \mathbf{u}) = \delta \widetilde{w}^{(c)}(\widetilde{\boldsymbol{\sigma}}) = \delta \widehat{w}^{(c)}(\widehat{\boldsymbol{\sigma}}) \tag{6b}$$

and

$$\hat{\boldsymbol{\sigma}} = \tilde{\boldsymbol{\sigma}} \quad \text{for } |\nabla_x \boldsymbol{u}| \ll 1 \text{ at (C)}. \tag{6c}$$

**M.2 Useful path integrals in incremental elasticity of affine hyperelastic finite deformation**

Now, we introduce path integrals $J_\alpha^{(\text{TM})}$ and $M^{(\text{TM})}$ in small ($|\nabla_x \boldsymbol{u}| \ll 1$) incremental-deformation fields of large hyperelastic 2D affine deformation, defined as

$$J_\alpha^{(\text{TM})} = \int_{\Gamma^{(\text{B})}} \{n_\alpha \hat{w}^{(c)}(\tilde{\boldsymbol{\sigma}}) - n_\gamma \tilde{\sigma}_{\beta\gamma} \hat{u}_{\beta,\alpha}\} ds \tag{7a}$$

$$M^{(\text{TM})} = \int_{\Gamma^{(\text{B})}} x_\alpha \{n_\alpha \hat{w}^{(c)}(\tilde{\boldsymbol{\sigma}}) - n_\gamma \tilde{\sigma}_{\beta\gamma} \hat{u}_{\beta,\alpha}\} ds \tag{7b}$$

where $\hat{u}_{\beta,\alpha} = S_{\beta\alpha\zeta\eta} \tilde{\sigma}_{\zeta\eta} = S_{\beta\alpha\zeta\eta} \sigma_{\zeta\eta}^{[B]} + u_{\beta,\alpha} + O(|\nabla_x \boldsymbol{u}|^2)$ with $\sigma_{\zeta\eta}^{[B]}$ being the Cauchy stress of the affine deformation and $\Gamma^{(\text{B})}$ is a contour in configuration (B). Since $\tilde{\boldsymbol{\sigma}}$ satisfies static equilibrium equations and incrementally linear elastic to $\nabla_x \hat{\boldsymbol{u}}$ and $\hat{w}^{(c)}(\tilde{\boldsymbol{\sigma}})$ is a quadratic functional of $\tilde{\boldsymbol{\sigma}}$ with a constant compliance tensor $\boldsymbol{S}$, $J_\alpha^{(\text{TM})}$ and $M^{(\text{TM})}$ become path independent for incrementally linear elastic fields of $|\nabla_x \boldsymbol{u}| \ll 1$, as shown in Appendix A. Since *bid* fields of surface flaws decay as $|\nabla_x \boldsymbol{u}| \sim 1/|\boldsymbol{x}|^2$, and $J_\alpha^{(\text{TM})}$ and $M^{(\text{TM})}$ are particularly useful for asymptotic analyses of the far field where $|\nabla_x \boldsymbol{u}| \to 0$. In addition, given that TM fields are incrementally linear elastic, other incrementally linear elastic fields of the same compliance are superposable, and interaction integrals of $J_\alpha^{(\text{TM})}$ and $M^{(\text{TM})}$, as bilinear-functionals of the two interacting fields, are also path independent.

The $J_\alpha^{(\text{TM})}$ integral is a subclass of the classical *J*-type path-independent integrals applicable to general nonlinear elastic deformation fields in a translationally homogeneous material. It is suitable for characterizing translationally self-similar field variations often encountered in studying defect motions (Eshelby, 1975; Kim et al., 2010) or fracture processes (Rice, 1968). On the other hand, the $M^{(\text{TM})}$ integral is a subclass of the classical *M*-type path-independent integrals applicable to linear-elastic deformation fields in radially homogeneous materials (Knowles &

Sternberg, 1972; Budiansky & Rice, 1973). It is convenient in describing expansionally or radially self-similar field variations often come across in examining cavity growth (Meyer et al., 2017), dislocation-core or -loop (Freund, 1978; Hurtado & Kim, 1999), or composite-wedge (Im & Kim, 2000) characterization, or volume-changing phase transformation (Markenscoff, 2021). In linear elastic systems, the interaction integrals (Chen & Shield, 1977; Hong & Kim, 2003) are applicable for both types of path-independent integrals. Since the TM incremental deformation over an affine nonlinear hyperelastic deformation is nominally linear elastic between the work conjugates, $\hat{\boldsymbol{\sigma}}$ and $\boldsymbol{\nabla}_x \hat{\boldsymbol{u}}$, both $J_\alpha^{(\text{TM})}$ and $M^{(\text{TM})}$ integrals are applicable to characterizing the TM fields of perturbed deformations made by surface flaws.

**M.3 Incremental elastic field of hyperelastic half-space under uniform lateral compression**

We will consider a surface-flaw field as an incremental elastic field of a traction-free hyperelastic half-space under uniform lateral plane-strain compression. While general incompressible hyperelastic half-space formulations are treated in Appendix B, we will first consider an incompressible neo-Hookean half-space in this section. The incompressible kinematics is depicted by a configurational mapping with a lateral stretch of $0 < \lambda \leq 1$ for compression as

$$x_1 = \lambda X_1; \quad x_2 = \lambda^{-1} X_2; \quad x_3 = X_3. \tag{8}$$

Then, the neo-Hookean constitutive relation, $\sigma_{ij} = \mu F_{ik} F_{jk} - p \delta_{ij}$, for $F_{ij}^{[0]} = \frac{\partial x_i}{\partial X_j}$ and $p$ being determined as $\mu/\lambda^2$ by traction-free boundary conditions, yields the fundamental stress state at (B),

$$\tilde{\sigma}_{11}^{(B)} = \sigma_{11}^{(B)} = \mu(\lambda^2 - \lambda^{-2}) \quad ; \quad \tilde{\sigma}_{33}^{(B)} = \sigma_{33}^{(B)} = \mu(1 - \lambda^{-2}) \quad ; \text{ all other } \tilde{\sigma}_{ij}^{(B)} = 0. \tag{9}$$

When a perturbation displacement field, $u_i(\boldsymbol{x}) = x_i^\dagger - x_i$, is imposed as an incremental elastic field, the incompressibility condition, $\|\mathbf{F}^{[1]}\| = 1$, is linearized for $|\boldsymbol{\nabla}_x \boldsymbol{u}| = |\mathbf{F}^{[1]} - \mathbf{I}| \ll 1$ as

$$u_{\alpha,\alpha} + O(|\boldsymbol{\nabla}_x \boldsymbol{u}|^2) = 0. \tag{10}$$

Once $\mathbf{F} = \mathbf{F}^{[1]}\mathbf{F}^{[0]}$ is inserted in the constitutive relation with $\mathbf{F}^{[1]} = \mathbf{I} + \nabla_x \mathbf{u}$ and ($F^{[0]}_{11} = \lambda$, $F^{[0]}_{22} = \lambda^{-1}$, $F^{[0]}_{33} = 1$, all other $F^{[0]}_{ij} = 0$), we get the Cauchy stress at (C) for $|\nabla_x \mathbf{u}| \ll 1$ as

$$\sigma_{11} = \mu\lambda^2 \left(1 + 2u_{1,1} + O(|\nabla_x \mathbf{u}|^2)\right) - p \tag{11a}$$

$$\sigma_{22} = \frac{\mu}{\lambda^2}\left(1 + 2u_{2,2} + O(|\nabla_x \mathbf{u}|^2)\right) - p \tag{11b}$$

$$\sigma_{12} = \sigma_{21} = \mu\left(\lambda^2 u_{2,1} + \frac{1}{\lambda^2} u_{1,2} + O(|\nabla_x \mathbf{u}|^2)\right). \tag{11c}$$

Introducing a stream function $\psi$: ($u_1 = \psi_{,2}$ and $u_2 = -\psi_{,1}$) and eliminating $p$ in (11a) and (11b) for static equilibrium, $\sigma_{\alpha\beta,\beta} = 0$, we find $\psi$ is duel-frame harmonic for $|\nabla_x \mathbf{u}||\nabla_X \nabla_x \mathbf{u}|/L \ll 1$,

$$\nabla_X^2 \nabla_x^2 \psi + O(|\nabla_x \mathbf{u}||\nabla_X \nabla_x \mathbf{u}|/L) = 0 \tag{12}$$

where $L$ is a characteristic length of the field. Solutions of (12) for general anisotropy can be expressed in two distinct analytic complex functions of $z = x_1 + ix_2$ and $Z = X_1 + iX_2$, respectively; however, the solutions become single-frame biharmonic for degenerative isotropy (Sun et al., 2011). Thereafter, we express the first-order perturbation displacement, $u_\alpha$, in terms of two analytical complex functions, $f(z)$ and $g(Z)$ as

$$u_1 = \text{Re}[f(z) + \lambda g(Z)] \tag{13a}$$

$$u_2 = -\text{Im}[f(z) + \lambda^{-1} g(Z)]. \tag{13b}$$

Then, the UPC-nominal stress, $\tilde{\boldsymbol{\sigma}} = \boldsymbol{\sigma}(\mathbf{I} + \nabla_x \mathbf{u})^{-T}$, is expressed from (4), (11), and (13) as

$$\tilde{\sigma}_{11} = \mu \frac{1}{\lambda^2} \text{Re}\left[2\frac{\partial f}{\partial z} + (1 + \lambda^4)\frac{\partial g}{\partial Z}\right] + \mu(\lambda^2 - \lambda^{-2}) \tag{14a}$$

$$\tilde{\sigma}_{22} = -\mu \frac{1}{\lambda^2} \text{Re}\left[(1 + \lambda^4)\frac{\partial f}{\partial z} + 2\frac{\partial g}{\partial Z}\right] \tag{14b}$$

$$\tilde{\sigma}_{12} = -\mu \frac{1}{\lambda^2} \text{Im}\left[2\frac{\partial f}{\partial z} + (\lambda^2 + \lambda^{-2})\frac{\partial g}{\partial Z}\right] \tag{14c}$$

$$\tilde{\sigma}_{21} = -\mu \frac{1}{\lambda^2} \text{Im}\left[(1 + \lambda^4)\frac{\partial f}{\partial z} + 2\lambda^2 \frac{\partial g}{\partial Z}\right]. \tag{14d}$$

Equations (13) and (14) are of the solution $\psi^{(0)} + \psi^{(1)}$ from a perturbation analysis with

$$\psi(\boldsymbol{x}; \boldsymbol{X}) = \psi^{(0)}(x_1, x_2; X_1, X_2) + \xi \psi^{(1)}(x_1, x_2; X_1, X_2) + \xi^2 \psi^{(2)}(x_1, x_2; X_1, X_2) + \cdots \tag{15}$$

that yields order-dependent governing equations with a perturbation parameter $\xi$ (Diab & Kim, 2014).

**M.4 Asymptotic far field of a surface flaw in a compressed hyperelastic half-space**

In this section, we consider admissible asymptotic far fields of incremental deformation in hyperelastic half-space under uniform lateral far-field compression, for which $f'(z)$ and $g'(Z)$ are analytic with $z$ and $Z$ respectively at $\infty$, vanish for $x_2 \to -\infty$, and make tractions free at $x_2 = 0$. The boundary conditions for $x_2 \to -\infty$ determine $p = \mu(\lambda^2 - \lambda^{-2})$ in (11). The traction-free boundary conditions at $x_2 = 0$ are expressed in terms of $f'(z)$ and $g'(Z)$ as

$$\text{Re}\left(\frac{(\lambda^2 + \lambda^{-2})}{2} f'(z) + \lambda^{-2} g'(Z)\right)_{x_2=0} = 0 \tag{16a}$$

$$\text{Im}\left(f'(z) + \frac{(\lambda^2 + \lambda^{-2})}{2} g'(Z)\right)_{x_2=0} = 0. \tag{16b}$$

The governing equation (12) and the boundary conditions at $x_2 = 0$ and $-\infty$ admit eigenfunction expansions of $f(z)$ and $g(Z)$.

One example is the Fourier eigenfunctions that depict a surface-wrinkling field, as expressed as

$$f(z) = A \exp(-\lambda^{-1} kzi) \tag{17a}$$

$$g(Z) = -\frac{(\lambda^4 + 1)a}{2\lambda} \exp(-kZi) \tag{17b}$$

for which $\lambda^4 + 1 - 2\lambda = 0$ must hold to satisfy the traction-free condition for arbitrary amplitude $A$ and wave number $k$, that identifies the Biot bifurcation point.

Another example is the power series expansion with singular Laurant eigenfunctions,

$$f(z) = \frac{2}{1-\lambda^4} \frac{B_0 i}{2\pi} \ln(z) + \sum_{n=1}^{\infty} a_n i^{n+1} z^{-n} \tag{18a}$$

$$g(Z) = -\frac{(1+\lambda^4)}{2\lambda} \frac{2}{1-\lambda^4} \frac{B_0 i}{2\pi} \ln(Z) + \sum_{n=1}^{\infty} b_n i^{n+1} Z^{-n} \tag{18b}$$

that satisfy the traction-free boundary conditions of (16). The boundary conditions require

$$a_n \frac{(\lambda^2 + \lambda^{-2})}{2} \lambda^{1-n} + b_n = 0 \quad \text{for odd } n \tag{19a}$$

$$a_n \lambda^{-1-n} + \frac{(\lambda^2 + \lambda^{-2})}{2} b_n = 0 \quad \text{for even } n \tag{19b}$$

and $B_0 = 0$, or $\lambda^4 + 1 - 2\lambda = 0$ for $B_0 \neq 0$ that spots the Biot bifurcation point. Here, $B_0$ implies the Burgers vector of a dislocation sitting at the origin of the $x$ coordinate. This result implies that $B_0$ must vanish unless $\lambda = \lambda_{cr}^{Biot}$, the root of $\lambda^4 + 1 - 2\lambda = 0$ ($\lambda \neq 1$). Thus, the incremental far-field stress of a surface flaw or a crease for $\lambda \neq \lambda_{cr}^{Biot}$ is represented by (18) with $B_0 = 0$, having leading-order terms of $n = 1$. The leading order terms exhibit $|\nabla u|^2 \sim |x|^{-4}$, making $O(|\nabla u|^2)$ in

(10) and (11) rapidly vanish for larger $|x|$, and the leading order field dominates the far field. Besides, the solution of a 2D point force applied at the origin also has a logarithmic form of $f(z)$ and $g(Z)$; however, the 2D point force violates the global force balance condition.

**M.5 TM configurational energetics of braced incremental deformation and its shape factor**

The singular Laurant series expansions (18) with $B_0 = 0$ correspond to the asymptotic far-field $\mathcal{L}(x) = \sum_{n=1}^{\infty} \mathcal{L}_n(x)$ of a surface flaw for $\lambda \neq \lambda_{cr}^{Biot}$, where $\mathcal{L}_n(x)$ represents a linear-elastic eigenfunction field of the $n_{th}$-power terms in (18). Taking the $M_{\Gamma_1+\Gamma_5+\Gamma_\infty}^{(TM)}[\mathcal{L}(x)]$ integral along a close contour in Fig. 2(a), the integral along $\Gamma_1$ and $\Gamma_5$ respectively vanishes, but $M_{\Gamma_\infty}^{(TM)}[\mathcal{L}(x)]$ does not. The nonvanishing $M_{\Gamma_\infty}^{(TM)}[\mathcal{L}(x)]$ implies that the linearized asymptotic far field has a leading order of $\sigma_{\alpha\beta} \sim 1/|x|^2$ since the nonvanishing $M_{\Gamma_\infty}^{(TM)}$ value only comes from the product of the leading order term in (18) with the uniform compression term at the far field. In addition, we can find an asymptotic far field of the same leading order that analytically continues to form a TM analytic-continuation-generated (ACG) near-field singularity of the *bid* field within the half-space for which $M_{\Gamma_3}^{(TM)}$ does not vanish. Configurational TM-ACG-singularity properties can be probed by interaction $\overleftrightarrow{M}_{\Gamma_\infty}^{(TM)}$ integrals between $\mathcal{L}(x)$ and various auxiliary conjugate fields $\mathcal{L}^{(ac)}(x)$;

$$\overleftrightarrow{M}_{\Gamma_\infty}^{(TM)}[\mathcal{L}(x); \mathcal{L}^{(ac)}(x)] = M_{\Gamma_\infty}^{(TM)}[\mathcal{L}(x) + \mathcal{L}^{(ac)}(x)] - M_{\Gamma_\infty}^{(TM)}[\mathcal{L}(x)] - M_{\Gamma_\infty}^{(TM)}[\mathcal{L}^{(ac)}(x)] \quad (20)$$

where $\Gamma_\infty$ is a contour in the far field shown in Fig 2(a) for $|x| \to \infty$. The $M_{\Gamma_\infty}^{(TM)}$ integral is defined in (7b) with nonzero TM compliance tensor components,

$$S_{1111} = S_{2222} = -S_{1122} = -S_{2211} = \frac{\lambda^2}{\mu(3+\lambda^4)} \quad (21a)$$

$$S_{1221} = S_{2112} = -S_{2121} = -\frac{S_{1212}}{\lambda^4} = \frac{\lambda^2}{\mu(1-\lambda^4)} \quad (21b)$$

for an incompressible 2D neo-Hookean half-space under uniform lateral compression.

Since the boundary conditions require the net traction force applied on the closed contour $\Gamma_1 + \Gamma_5 + \Gamma_\infty$ to vanish, the TM-ACG singularity must be a self-equilibrium singularity of $\sigma_{\alpha\beta} \sim 1/r$ at the near field for nonvanishing $M_{\Gamma_3}^{(TM)}$. In addition, the configurational image force of the traction-free boundary exerted on the singularity must balance the configuration force applied on the singularity by the lateral compression, as shown in Fig. 2(b). Such singularity is a TM-ACG subsurface dislocation of a Burgers vector of magnitude $B$ sitting beneath the traction-free surface in depth $h$, as shown in Fig. 2(c). The analytic functions for the TM-ACG subsurface dislocation field are derived as

$$f^{(ssd)}(z) = \frac{iB}{\pi\lambda(1-\lambda^4)}\left\{\lambda \ln(z+hi) - \frac{\lambda a}{b}\ln(z-hi) + \frac{2\lambda c}{b}\ln(z-\lambda^2 hi)\right\} \quad (22a)$$

$$g^{(ssd)}(Z) = \frac{-iB}{\pi\lambda(1-\lambda^4)}\left\{\lambda c \ln(Z+\lambda hi) + \frac{\lambda ac}{b}\ln(Z-\lambda hi) - \frac{2\lambda}{b}\ln(Z-\lambda^{-1}hi)\right\} \quad (22b)$$

where $a = c^{-1} + c$, $b = c^{-1} - c$ with $c = (1+\lambda^4)/(2\lambda)$. The subsurface dislocation quantities are measured in $(x_1, x_2)$ frame. It is noted that $b$ vanishes at the Biot critical point, which makes $f(z)$ and $g(Z)$ singular with respect to $\lambda$. Equations (22a&b) represent a superposition of the fields of a TM-ACG dislocation sitting at $x_2 = -h$ and three image dislocations located at $x_2 = \lambda^2 h, h,$ and $\lambda^{-2}h$ in the entire domain that nullify the traction at $x_2 = 0$, as shown in Fig 2(c). Now, as we have the TM-ACG subsurface-dislocation TM field at hand with (22) in Appendix C, we can characterize the *bid* field of a surface flaw with the Burgers vector $B$ and the location $h$ of the TM-ACG subsurface dislocation, evaluating $M_{\Gamma_\infty}^{(TM)}[\mathcal{L}(x)]$ and $\overleftrightarrow{M}_{\Gamma_\infty}^{(TM)}[\mathcal{L}(x);\mathcal{L}^{(ac)}(x)]$ as follows.

As $M_\Gamma^{(TM)}$ integration along a close contour $\Gamma = \Gamma_\infty + \sum_{k=1}^{5}\Gamma_{\infty k}$ in Fig. 2(a) vanishes, we have

$$M_{\Gamma_\infty}^{(TM)}[\mathcal{L}(x)] = -M_{\Gamma_3}^{(TM)} \equiv M_{core}^{(TM)} = \frac{B^2}{4\pi}H\left(\mathcal{S}_{\alpha\beta\zeta\eta}(\lambda)\right) \quad (23a)$$

or

$$B = \left\{ \frac{4\pi M_{\Gamma_\infty}^{(TM)}[\mathcal{L}(\boldsymbol{x})]}{H\left(\mathcal{S}_{\alpha\beta\zeta\eta}(\lambda)\right)} \right\}^{\frac{1}{2}} \tag{23b}$$

where $M_{\Gamma_3}^{(TM)} = -hJ_{dslcn}^{(TM)} - M_{core}^{(TM)}$ in which $J_{dslcn}^{(TM)} = J_{image}^{(TM)} - J_{\tilde{\sigma}_{11}^\infty}^{(TM)} = 0$ for configurational force balance. The stiffness pre-factor $H\left(\mathcal{S}_{\alpha\beta\zeta\eta}(\lambda)\right)$ of a dislocation core in an anisotropic linear elastic TM field of a general incompressible hyperelastic solid is presented in (B15) of Appendix B. In addition, an $\overleftrightarrow{M}_{\Gamma_\infty}^{(TM)}$ integral for interaction between the field $\mathcal{L}(\boldsymbol{x})$ and an auxiliary field $\mathcal{L}^{(\tilde{\sigma}_{11}^0)}(\boldsymbol{x})$ of an additional uniform compression $\tilde{\sigma}_{11}^0$ leads to

$$\overleftrightarrow{M}_{\Gamma_\infty}^{(TM)}\left[\mathcal{L}(\boldsymbol{x}); \mathcal{L}^{(\tilde{\sigma}_{11}^0)}(\boldsymbol{x})\right] = \tilde{\sigma}_{11}^0 Bh \tag{24a}$$

or, by inserting (23b) for B,

$$h = \frac{\overleftrightarrow{M}_{\Gamma_\infty}^{(TM)}\left[\mathcal{L}(\boldsymbol{x}); \mathcal{L}^{(\tilde{\sigma}_{11}^0)}(\boldsymbol{x})\right]}{\tilde{\sigma}_{11}^0} \left\{ \frac{4\pi M_{\Gamma_\infty}^{(TM)}[\mathcal{L}(\boldsymbol{x})]}{H\left(\mathcal{S}_{\alpha\beta\zeta\eta}(\lambda)\right)} \right\}^{-\frac{1}{2}}. \tag{24b}$$

Equations (23) and (24) set up a definition of *bid*-field shape factor

$$S^{(bid)}(\lambda) \equiv \frac{B}{4\pi h} = \frac{\tilde{\sigma}_{11}^0 M_{\Gamma_\infty}^{(TM)}[\mathcal{L}(\boldsymbol{x})]}{H\left(\mathcal{S}_{\alpha\beta\zeta\eta}(\lambda)\right) \overleftrightarrow{M}_{\Gamma_\infty}^{(TM)}\left[\mathcal{L}(\boldsymbol{x}); \mathcal{L}^{(\tilde{\sigma}_{11}^0)}(\boldsymbol{x})\right]} \tag{25}$$

for a *bid* field of a surface flaw, which is in configurational equilibrium in a hyperelastic half-space compressed by a lateral loading. Here, we consider a *bid* field generated by a surface flaw such as a notch, a surface pinching, a crease, or a nonlinear deformation field of a 2D point load subtracted by a linear-elastic field of the 2D point load.

The *bid*-field shape factor is explicitly expressed for neo-Hookean solids as

$$S_{\text{nH}}^{(bid)}(\lambda) = \left(\frac{B}{4\pi h}\right)_{\text{nH}} = \frac{(1-\lambda^2)(\lambda+\lambda^3)^2}{-1+3\lambda^2+\lambda^4+\lambda^6}, \tag{26a}$$

once we insert into (25)

$$H_{nH}\left(S_{\alpha\beta\zeta\eta}(\lambda)\right) = \mu\left(\frac{-1+3\lambda^2+\lambda^4+\lambda^6}{\lambda^4+\lambda^6}\right), \tag{26b}$$

$$M_{\Gamma_\infty}^{(\text{TM})}[\mathcal{L}_{nH}(\boldsymbol{x})] = \mu B h(\lambda^2-\lambda^{-2}), \tag{26c}$$

and

$$\overleftrightarrow{M}_{\Gamma_\infty}^{(\text{TM})}\left[\mathcal{L}_{nH}(\boldsymbol{x}); \mathcal{L}_{nH}^{(\tilde{\sigma}_{11}^0)}(\boldsymbol{x})\right] = \tilde{\sigma}_{11}^0 Bh, \tag{26d}$$

where the subscript '*nH*' implies neo-Hookean. We derived the $H_{nH}\left(S_{\alpha\beta\zeta\eta}(\lambda)\right)$ expression in (26b) by directly integrating the dislocation field articulated in (18) to get $M_{core}^{(\text{TM})}$. The neo-Hookean *bid*-field shape factor $S_{\text{nH}}^{(bid)}$ in (26) is a monotonically increasing function of the far-field compressive strain $(1-\lambda)$, as shown in Fig. 3(d). The *bid*-field shape factor $S_{\text{nH}}^{(bid)}$ vanishes at the null strain of the far field and increases with the far-field compression until the substrate creases when $S_{\text{nH}}^{(bid)}$ reaches the *crease-limit shape factor* $S_{\text{nH}}^{(cr)}$. What sets the shape factor that only depends on $\lambda$? The free-surface boundary condition imposed on the *bid* field develops the shape factor with which a self-similar *bid* field is in configurational equilibrium for single-parameter, *B*, variations for every fixed $\lambda$. If the configurational equilibrium of a *bid* field is stable, the shape factor represents the nonlinear eigenmode of the *bid* field, whose magnitude is determined by the flaw size. If unstable, the shape factor corresponds to a self-similar growth eigenmode of creasing. In the following sections, we test the stability of the configurational equilibrium for the self-similar *bid*-field variations at every fixed $\lambda$ to find the crease limit point at which a singular perturbation with a surface flaw triggers configurational instability of the self-similar *bid* field to develop creasing. Here, we name the higher-order instability point 'crease limit point' to distinguish the higher-order instability point from the 'wrinkle critical point' of Biot's first-order collapse instability of the surface flatness.

**Computational Analysis**

The first-order singular perturbation can only address the configurational equilibrium of the self-similar *bid* field, and the stability of the equilibrium requires higher-order perturbation analysis in (15) or finite element (FEM) analysis of the *bid* field. Here, we test the stability with finite element analysis since we have developed computational tools, $M_{\Gamma_\infty}^{(TM)}[\mathcal{L}(x)]$ and $\overleftrightarrow{M}_{\Gamma_\infty}^{(TM)}[\mathcal{L}(x); \mathcal{L}^{(\tilde{\sigma}_{11}^0)}(x)]$ formulas in (23a) and (24a) for the stability test. The FEM analysis also reveals strain energy density distribution near the *bid* field at various compression stages. At the crease limit point where the higher-order configurational instability sets in, we will match the near and the far field asymptotes as we have a single-parameter scaling solution, each for both interior and exterior asymptotic fields. In addition, since we have developed mathematical tools to obtain asymptotic scaling solutions of creasing in general incompressible hyperelastic solids, we also investigate material-property effects on the crease limit point with the Gent model (1996) in this section.

**C.1 FEM simulations for crease limit-point identification in hyperelastic solids**

Figure 3(a) shows a schematic of the FEM analysis domain ($100 \times 100\ unit$) that includes a *bid* field and a range ($l = 37, t = 8\ unit$ in Fig. 3(a)) of $M^{(TM)}$ domain integral. The $unit$ is a reference length in the undeformed configuration (A) in Fig. 1(b) that deforms with the body to the configuration (B). For FEM simulations, we employed ABABUS/Standard with stabilized Newton–Raphson method and a plane-strain bilinear reduced-integration hybrid element, CPE4RH, adopting logarithmic mesh spacings with the inner element size of $0.01\ unit$ and outer element size of $1.0\ unit$. An incompressible neo-Hookean constitutive relation was used in the simulations. All $M^{(TM)}$ integrals were computed in the $(x_1, x_2)$ coordinate of the configuration (B). We employed a domain integral scheme (Nakamura et al., 1985), ensuring the computational accuracy and path independence of $M^{(TM)}$ integrations. The upper insets also displayed four different perturbation-flaw configurations to generate various *bid* fields. The surface-flaw configurations include a surface cusp of depth *d* made of two quarter-circles or two quarter-ellipses, a wedge-shaped notch of depth *d*, or a 2D point load *T*. All tested surface flaw depths were $d(= 0.2\ unit)$.

In order to simulate the half-space deformation using a finite-size FEM-simulation domain, we developed an iterative method of matching the FEM-simulation near-field and the theoretical asymptotic far-field. In the first step, mixed boundary conditions are imposed to have the top surface traction free, the two side boundaries respectively move $\pm\Delta(1-\lambda)\ unit$ at the left and right horizontal directions with $\tilde{\sigma}_{12} = 0$, and the bottom boundary holds no normal displacement and no shear traction. From the first step computation, we evaluated the first iteration values of the TM-ACG dislocation Burgers vector $B_{(1)}$ and depth $h_{(1)}$ with (23b) and (24b) by computing $M_{\Gamma_\infty}^{(TM)}[\mathcal{L}(x)]$ and $\overleftrightarrow{M}_{\Gamma_\infty}^{(TM)}[\mathcal{L}(x); \mathcal{L}^{(\tilde{\sigma}_{11}^0)}(x)]/\tilde{\sigma}_{11}^0$. Then, we apply the theoretical asymptotic-field displacements of $B_{(1)}$ and $h_{(1)}$ in (13) and (22) as boundary conditions of the first iteration FEM simulation. From the first iteration FEM simulation, we compute $B_{(2)}$ and $h_{(2)}$ for the second iteration that will be used to impose the second-iteration boundary displacements. These iterations are repeated $n$ times until $B_{(n)}$ and $h_{(n)}$ converge to stationary values. Figure 3(b) shows the convergence of $B_{(n)}$, $h_{(n)}$, and $S_{(n)}^{(bid)} (= B_{(n)}/4\pi h_{(n)})$ in an iterative half-space analysis with finite-domain simulations for $1 - \lambda = 0.344$. Variations of $B_{(n)}$ and $h_{(n)}$ values are approximately 1% off from the initial estimates with the finite domain FEM simulation to the half-space values, while the variation in $S_{(n)}^{(bid)}$ is negligible within 0.05%. The variations increase, and the convergence speed slows down when $S_{(n)}^{(bid)}$ approaches $S_{(n)}^{(cr)}$ of the crease limit point for which $1 - \lambda \approx 0.356$. Near the crease limit point, $B$ and $h$ are observed to be proportional to the flaw-tip indentation displacement, $\Delta_{tip}$, as $B \approx 1.82\Delta_{tip}$ and $h \approx 0.145\Delta_{tip}$, indicating that the shape factor is fixed during the crease-tip advancement with only one length scale varying.

Figure 3(c) displays FEM-evaluated $M_{\Gamma_\infty}^{(TM)}[\mathcal{L}(x)]/\mu d^2$ variations as functions of $1 - \lambda$ for compressive loading and unloading. It is reversible except for a slight jump close to the crease point, at $1 - \lambda \approx 0.355$, during the loading process, possibly due to computational error fluctuation. However, the fluctuation is not observed in the unloading curve; instead, a tiny fluctuation occurs near $1 - \lambda \approx 0.351$ while unloading. By and large, the $M_{\Gamma_\infty}^{(TM)}[\mathcal{L}(x)]/\mu d^2$ variation is negligible below $1 - \lambda = 0.350$ but nearly linearly varies in the range of $0.350 < 1 - \lambda = 0.360$ across the crease point. In contrast, the *bid*-field shape factor, $S^{(bid)}(= B/4\pi h)$, monotonically increases from zero until it reaches $S^{(bid)} = S^{(cr)}$, where it creases near $1 - \lambda \approx$

0.355, as shown in Fig. 3(d). The crease-limit shape factor is identified as $S^{(cr)} = 1$ within the computational accuracy. Then, the crease-limit stretch ratio becomes $\sqrt{\sqrt{2}-1} \cong 0.644$, the positive real root of $\lambda^4 + 2\lambda^2 - 1 = 0$ from (26a). The *bid*-field shape factor, $S^{(bid)}$, was also computationally evaluated with FEM simulations (25) and (26b) for various surface flaws, including a 2D point load $T$, and displayed in Fig. 3(d). The computational *bid*-field shape factor closely follows the theoretical curve until it creases, regardless of the flaw shapes. We also observed that the *bid*-field shape factor is indifferent to different flaw sizes and $T$ values. To evaluate the *bid*-field shape factor of the 2D point load with FEM analysis, we used the following formula;

$$S_{nH}^{(bid)}(\lambda) = \frac{\left(M_{\Gamma_\infty}^{(TM)}[\mathcal{L}_{nH}(x)] - M_{Tcore}^{(TM)}\right)\tilde{\sigma}_{11}^0}{H_{nH}\left(\mathcal{S}_{\alpha\beta\zeta\eta}(\lambda)\right)\vec{M}_{\Gamma_\infty}^{(TM)}\left[\mathcal{L}_{nH}(x); \mathcal{L}_{nH}^{(\tilde{\sigma}_{11}^0)}(x)\right]} \tag{27a}$$

with the $M_\Gamma^{(TM)}$-integral around the 2D point load core,

$$M_{Tcore}^{(TM)} = -\frac{1}{1+\lambda^2}\frac{T^2}{4\pi\mu}. \tag{27b}$$

To reveal how the shape factor determines the crease-stability limit point, we plotted the FEM-simulated strain-energy density distributions near a surface flaw at four different compression levels of $1-\lambda = 0.1, 0.2, 0.3$, and $0.344$ in Fig. 4(a-d). Figure 4(e) displays the energy distribution in the TM-ACG subsurface-dislocation field of the same compression level as (d). Comparison between (d) and (e) shows the TM-ACG field is already close to the fully nonlinear deformation field within a two-Burger-vector distance from the tip. All the frames of (a)-(e) in Fig. 4 imply that the leading-order primary mode of incremental elasticity, $f(z) = -a_1 z^{-1}$ and $g(Z) = -b_1 Z^{-1}$ in (18a&b), distinctly exhibits three-fold energy-release angular sectors separated by two energy-elevation sectors at the far field. The light yellowish energy-release sectors have lower energy density than the uniform lateral-compression energy density at the far field. In contrast, the darker brownish-color energy-elevation sectors have higher energy density. On the other hand, the close-tip field has a single-fold strain energy distribution so that a zone of mode transition from a three-fold symmetry to a single-fold symmetry exists between the far and near fields. The *bid*-field shape factor, $B/4\pi h$, is believed to characterize the mode-transition zone configuration. The neck of the energy-release domain in the mode transition zone broadens. At the same time, the two tips of the energy-elevation sectors move close to the free surface as the lateral compression increases from (a) to (d) in Fig. 4. Overall, Three radial modal zones and three angular energy-release sectors constitute the incremental deformation field caused by a surface flaw in the hyperelastic half space under lateral compression. The FEM analysis shows that the *bid* field is configurationally stable for self-similar expansion variations until the shape factor reaches the crease-limit value. At the crease-limit point, the stability of the configuration is neutral for a single length parameter, $B$, self-similar expansion variation of the deformation field.

### C.2 Matched Asymptotes of creasing in an incompressible neo-Hookean half-space

Once creasing starts, self-contact folding kinematics of creasing (Silling, 1991),

$$r = \frac{R}{\sqrt{2}} \tag{28a}$$

$$\theta = 2\Theta + \frac{\pi}{2} \quad \text{for } -\pi \leq \Theta \leq 0 \tag{28b}$$

generates an asymptotic near-field deformation of radial stretch $1/\sqrt{2}$ and angular stretch $\sqrt{2}$ for $r = |x^\dagger| \to 0$, where $R = \sqrt{X_1^2 + X_2^2}$, $\Theta = \tan^{-1}(X_2/X_1)$, $r = \sqrt{x_1^{\dagger 2} + x_2^{\dagger 2}}$, $\theta = \tan^{-1}(x_2^\dagger/x_1^\dagger)$. The equilibrium equation of neo-Hookean solids for the self-contact crease folding kinematics (28a&b) yields the pressure distribution,

$$p = -\frac{3}{2}\log\left(\frac{R}{R_c}\right) + 2 \tag{28c}$$

for which $R = R_c$ denotes the fold-contact zone end location where the normal traction vanishes. Then, the strain energy density, $(\mu/2)(\sum_{i=1}^{3} \lambda_i^2 - 3)$, uniformly converges to $\mu/4$ for $-3\pi/2 \leq$

$\theta \leq \pi/2$, as $r \to 0$. The self-contact folding kinematics generates a radially self-similar eigenmode of an asymptotic crease-tip deformation field with an undetermined amplitude $R_c$.

At the far field, an asymptotic incremental-deformation field develops with an undetermined amplitude $B$. The asymptotic far field is denoted by a radially self-similar duel-frame harmonic eigenmode of an incremental deformation set up by the global force balance and the traction-free surface boundary conditions. We designate superscripts, $\pm$, for the far and the near asymptotic fields in the following. Figure 5(a) displays a convergence of the incremental UPC-nominal stress, $\tilde{\sigma}_{11}^+(0,x_2) - \tilde{\sigma}_{11}^\infty$, to the leading-order field of $\sigma \sim 1/|x|^2$ (a dashed line) at the far field for $1 \ll |x_2|/B$, as obtained by iterative FEM simulations (solid lines) and a TM-ACG subsurface-dislocation model analysis (a dash-dot line). The stress field converges fast to the leading-order stress field for $25 < |x_2|/B$. In addition, the iterative FEM simulations converged to the leading order field in three iterations. Figure 5(b) shows the incremental UPC-nominal stress distributions, $\tilde{\sigma}_{11}^-(0,x_2) - \tilde{\sigma}_{11}^\infty$, for $0 \leq |x_2|/B \leq 5$. Both the incremental stress distributions in the near (a dotted line) and the far field asymptotic solutions exhibit sign change along $x_2$ from compression to tension as $|x_2|$ increases. The sign change is caused by the free-traction boundary condition imposed on the near-tip deformation field generated by the self-contact folding kinematics of the crease tip. The two fields, respectively, have a single scaling length parameter, $R_c$ and $B$. Here, the two parameters are to be reduced to a single scaling parameter by matching the asymptotes as follows.

The incremental UPC-nominal stress, $\tilde{\sigma}_{11}(0,x_2) - \mu(\lambda^2 - \lambda^{-2})$, switches the sign at

$$x_2^\circ|_{exterior} = -\frac{\lambda^2(3+\lambda^4)}{(1-\lambda^4)}h \qquad (29a)$$

in the exterior TM-field solution, and at

$$x_2^\circ|_{interior} = -\frac{R_c}{\lambda}\exp\left\{\frac{2\sqrt{2}(\lambda^{-2}-\lambda^2)}{3\lambda}\right\} \qquad (29b)$$

in the interior crease-tip solution. By matching the sign-switching location, $x_2^\circ$, of the incremental UPC-nominal stress, we have the interior and the exterior UPC-nominal stresses, $\tilde{\sigma}_{11}^\pm(0, x_2^\circ)$, coincide with the uniform compression, $\mu(\lambda^2 - \lambda^{-2})$, yielding,

$$\frac{R_c}{B} = \frac{\lambda^3(3 + \lambda^4)}{4\pi(1 - \lambda^4)} \exp\left(\frac{2\sqrt{2}(\lambda^{-2} - \lambda^2)}{3\lambda}\right) \approx 1.52 \tag{30a}$$

$$\frac{x_2^\circ}{B} = -\frac{\lambda^2(3 + \lambda^4)}{4\pi(1 - \lambda^4)} \approx -0.126 \tag{30b}$$

with $B/4\pi h = 1$ and $\lambda = \sqrt{\sqrt{2} - 1} \approx 0.644$ for the creasing configuration. Figure 5(b) displays the matched asymptotic solutions of the crease-tip and the subsurface-dislocation, comparing with an FEM simulation result of $1 - \lambda = 0.353$. As $1 - \lambda$ approaches the theoretical crease-limit value of 0.356, the crease configuration becomes too compliant for the iterative FEM half-space simulation to reach the final stationary configuration fast enough. At $1 - \lambda = 0.353$, the scaling ratio, $R_c/B$, reached 1.58.

### C.3 Strain-stiffening effect on hyperelastic creasing: The Gent-model

Mooney (1943) and Rivlin (1948) introduced a more general strain-energy density function $w(\lambda_1, \lambda_2, \lambda_3)$ of isotropic incompressible hyperelastic solids than that of neo-Hookean solids as

$$w(\lambda_1, \lambda_2, \lambda_3) = \frac{\mu}{2}(I_1 - 3) + C_{02}(I_2 - I_1); \quad \lambda_1\lambda_2\lambda_3 = 1 \tag{31}$$

where $\lambda_1, \lambda_2, \lambda_3$ are principal stretches. Here, $\mu$ is the linear-elastic shear modulus at small strain for isotropic variations of $I_1 = \lambda_1^2 + \lambda_2^2 + \lambda_3^2$, and $C_{02}$ is another stiffness to adapt additional isotropic variations of $I_2 = \lambda_1^2\lambda_2^2 + \lambda_2^2\lambda_3^2 + \lambda_3^2\lambda_1^2$. However, plane-strain deformations of $\lambda_3 = 1$ for $\lambda_1\lambda_2\lambda_3 = 1$ yield $I_2 - I_1 = 0$, and the Mooney-Rivlin solid is reduced to a neo-Hookean solid for plane strain deformations, regardless of $C_{02}$. Therefore, the crease characteristics of incompressible Mooney-Rivlin solids are identical to those of incompressible neo-Hookean solids.

Besides the Mooney-Rivlin model, Gent (1996) introduced another isotropic incompressible hyperelastic model that well describes the behavior of strain-stiffening crosslinked rubbery materials with

$$w(I_1; \mu, J_m) = -\mu \frac{J_m}{2} \ln\left(1 - \frac{I_1 - 3}{J_m}\right); \quad \lambda_1 \lambda_2 \lambda_3 = 1 \tag{32a}$$

where $J_m$ is a material constant that limits stretches within $I_1 - 3 < J_m$. The Gent model converges to the neo-Hookean model for $(I_1 - 3)/J_m \ll 1$, and the stiffness of distortional deformation diverges as the distortional deformation $I_1 - 3$ approaches $J_m$. The stretch limit of $I_1 - 3 < J_m$ is reduced to the Gent limit of compression for a plane-strain compression $\lambda$ of principal stretches $(\lambda, \lambda^{-1}, 1)$, as

$$\frac{1}{2}\left[J_m + 2 - \sqrt{J_m(J_m + 4)}\right] < \lambda \leq 1. \tag{32b}$$

The Gent limit of compression is plotted in Fig. 6(b) with a thin dashed line.

Chen et al. (2014) investigated the crease-limit instability of the Gent model with FEM analyses and reported the crease-limit compression stretch ratio $\lambda^{(\text{cr})}$ as a function of the strain-stiffening parameter $J_m$. They found that the function $\lambda^{(\text{cr})}(J_m)$ has two branches, $\lambda^{(\text{cr})+}(J_m)$ and $\lambda^{(\text{cr})-}(J_m)$, as displayed with black solid circles in Fig. 6(b). Branch $\lambda^{(\text{cr})+}(J_m)$ represents the crease limit of *bid* fields for the appearance of creasing upon compression, and branch $\lambda^{(\text{cr})-}(J_m)$ corresponds to the disappearance of the crease upon further compression. We noticed that the crease-limit shape factor $S^{(\text{cr})}$ depends on $J_m$, and we best fitted $S^{(\text{cr})}(J_m)$ for three parameters, a, b, and c, with the data $\lambda^{(\text{cr})+}(J_m)$ as

$$S^{(cr)}(J_m) = (1 - \breve{a})\sqrt{\left(1 - e^{-\breve{b}(J_m - \breve{c})}\right)} + \breve{a} \tag{32c}$$

where $\breve{a} = 0.611$, $\breve{b} = 0.104$, and $\breve{c} = 3.26$ as shown in Fig. 6(a). Then, branch $\lambda^{(cr)-}(J_m)$ could be predicted as shown with open circles in Fig. 6(b), which also exhibits $\lambda^{(cr)\pm}(J_m)$ curves for various fixed $S^{(cr)}$ values. The result (32c) implies that strain stiffening with stretch limits, (32b), improves crease resistance of the *bid* fields, resulting in no crease for $J_m < 3.26$.

## Discussion

Considering the far field perturbed by a surface flaw over a uniform lateral-compression field of a hyperelastic half-space, the perturbed incremental displacement gradient is a work conjugate of the nominal stress on the unperturbed affine-deformation configuration. The nominal stress and the incremental displacement gradient constitute an incremental linear elasticity in every tangential manifold of the hyperelastic affine finite deformation. The incremental linear elasticity supports path independences of $\boldsymbol{J}^{(TM)}$ and $M^{(TM)}$ integrals in the TM fields. Furthermore, the nominal stress and the TM displacement fields can be expressed with duel-frame harmonic functions represented by two distinct analytic functions of two complex variables, $z\ (= x_1 + ix_2)$ and $Z\ (= X_1 + iX_2)$, respectively, for the anisotropic incremental responses. The complex functions of the perturbed far field have leading-order stress singularities of $1/z^2$ and $1/Z^2$, respectively, in their Laurent series that represent the singular-perturbation field. The leading-order singularities interact with uniform lateral compression to yield non-vanishing integral values of $M_{\Gamma_\infty}^{(TM)}$ and $\overleftrightarrow{M}_{\Gamma_\infty}^{(TM)}(\tilde{\sigma}_{11}^0)$. There is a unique TM-ACG subsurface dislocation with the same leading-order singularity at the far field, and the two $M$ integrals at the far field determine the Burgers vector $B$ and the subsurface depth $h$ of the dislocation. The far-field braces the surface flaw's energy-releasing near-tip field against growing by forming a *bid* field, and the far field of the TM-ACG subsurface dislocation represents that of the *bid* field.

In previous sections, we found that the shape factor, $S^{(bid)}(\lambda) = B/4\pi h$, determines the *bid* field's configurational stability against self-similar growth. The *bid* field is in configurational equilibrium until the far-field compression increases to have $S^{(bid)}$ reach the crease-limit shape factor $S^{(cr)}$. The far-field compression strengthens both the flaw-tip field's energy release mode and the incremental far field's bracing mode. However, incremental hyperelasticity's anisotropy tends to weaken the far field's bracing capacity to activate the higher-order crease instability at the crease limit point. Nevertheless, these analyses are based on 2D singular perturbation of a fundamental state of uniform compression. If a 3D flaw in a non-uniform compression field is considered, the crease-limit strain values can vary from the 2D prediction of the fundamental-state stability. In addition, the imperfection sensitivity of the crease limit point depends on the surface waviness, the material's microstructural inhomogeneity, compressibility, surface energy, finite specimen dimension, and the detection accuracy of the limit point. These imperfection sensitivity sources are believed the cause of the wide range of the scattered experimental measurement values of the limit point, $1 - \lambda^{(cr)} = 0.35 \pm 0.07$, for a rubber (Cho & Gent, 1999). On the other hand, recent experiments on Hydrogels and PDMS showed narrower scatter bands of the crease-limit strains with mean values within $0.35 - 0.36$ (e.g., Trujillo et al., 2008; Jin et al., 2021) which were close to the 2D prediction.

The imperfection sensitivity of surface waviness on the local crease-limit threshold was investigated with supporting FEM analyses (Cao & Hutchinson, 2012) and of the graded-stiffness boundary layer with Koiter's (1945) higher-order stability analysis that revealed the setback-crease mechanism (Diab et al., 2013; Diab & Kim, 2014). These analyses showed that local creases prematurely nucleated by local strain concentration could be confined in a boundary layer if the global 2D crease limit is not reached. Generally, a stiff surface layer causes wrinkling instability (Shield et al., 1994; Sun et al., 2011) below the crease limit and can trigger premature local creasing instability. On the other hand, a compliant pretension layer can protect the flatness of the substrate, delaying crease formation up to the flatness collapse point like Biot's critical point of neo-Hookean solids. An extreme example of a pretension layer is surface tension. The surface tension ought to delay crease nucleation from a surface flaw. Once creased, the self-contact size is no longer an undetermined parameter and will jump to an equilibrium length set by the surface energy. Unzipping the crease, the contact length vs. strain relation would follow a smooth unloading path. Indeed, elastocapillary (Liu et al., 2019) and adhesive-surface (Van Limbeek et al., 2021) creases exhibit surface-flaw shape sensitivity of the crease-folding memory on the crease injection and ejection hysteresis. If a crease is created without surface tension at the surface-flatness collapse point like the Biot critical point, the crease tip would dynamically jump to an equilibrium configuration in a nonuniform nonlinear deformation field of the external loading like a bending field (Hohlfeld & Mahadevan, 2011).

We introduced the shape factor of the *bid* field as a measure of the *bid* field's closeness to the crease limit, where the crease field self-similarly expands with a single undetermined scaling length parameter. At this point, the interior field scaling parameter, the crease-tip contact length, could be unified with the exterior field scaling parameter, the Burgers vector of the TM-ACG subsurface dislocation, through matched asymptotes. The newly introduced crease-limit shape factor has an analogous character in hyperelastic creasing to the stress intensity factor's (Williams, 1957) role in the Griffith (1921) fracture criteria. A difference between the two is that the crease-limit shape factor is for self-similar expansion, while the stress intensity factor is for self-similar translation of singular fields. The crease-limit shape factor and the TM-field energetics introduced in this paper will be particularly useful in studying the strength and durability of soft materials' surfaces and interfaces in various chemical or bio environments. Examples include stress-gradient coupled failure processes associated with crease tip singularities in inhomogeneous large deformations, like diffusion-coupled or radiation-induced oxidation or fatigue damage of soft materials with pinning scars under repeated loading (Van Limbeek et al., 2021). The TM energetic formalism will also be valuable for exploring ruga mechanics of self-organization (Zhao et al., 2015 & 2016) and designing programmable matter (e.g., Hawkes et al., 2010) and skin conditions (e.g., Matsumoto et al., 2010) for various applications. The incremental hyperelasticity formalisms in this study will help develop quantum devices with graphene crinkles (Kothari et al., 2018 & 2019; Kothari & Kim, 2022), soft robotics (e.g., Rus & Tolley, 2015; Psarra et al., 2019; Ze et al., 2020) or smart textiles (Persson et al., 2018; Adak & Mukhopadhyay, 2023).

**Conclusion**

In this work, we have found that the crease-tip motion is a configurational variation of a self-similar singular-field expansion in contrast to a self-similar singular-field translation in a fracture process. The crease-tip's self-similar-expansion singular field is composed of respectively self-similar interior and exterior expansion singular fields. The findings were made possible by carefully sorting hyperelastic incremental deformation and singular perturbation analyses with duel-frame harmonic functions for the exterior singular field.

Our new analytic solutions of the exterior asymptotic field allowed us to develop an iterative FEM analysis of the half-space incremental deformation with finite-domain simulations. Matched asymptotes could unify two scaling length parameters of the interior and exterior singular fields. Our FEM-simulation results converged to the two asymptotic fields. The new iterative FEM algorithm of the half-space analysis accurately detects the compression threshold of the crease-limit transition point for a surface flaw's singular perturbation field.

A near-tip energy-releasing field of a surface flaw is the interior singular field of a single angular mode which tends to transition to a crease singular field under far-field compression. The exterior singular field is composed of three energy-release angular sectors separated by two energy elevation sectors, which tend to brace the interior field against creasing. The conflict between the interior and the exterior singular fields constructs a *bid* field that keeps the stability of the field against transition below the crease-limit compression. We identified the *bid*-field shape factor that determines the degree of the flaw-tip field stability against creasing.

Two path-independent $M^{(\text{TM})}$ integrals determine the bid-field shape factor at the far field, which allowed us to find the TM-ACG subsurface dislocation. The bid-field shape factor is the ratio of the Burgers vector to the subsurface depth of the dislocation. The *bid*-field shape factor is smaller than unity below crease-limit compression in neo-Hookean solids; the surface is stable for both regular and singular perturbations below the crease-limit compression of 0.356 in neo-Hookean solids. At the crease-limit point, the shape factor becomes the crease-limit shape factor of unity, where the *bid* field undergoes a higher-order transition to a neutrally unstable crease field. The flat surface is metastable between the crease limit compression of 0.356 and the wrinkle collapse compression of 0.456 for neo-Hookean solids; it is stable for regular perturbations but unstable for singular perturbations.

We made the singular perturbation analyses for neo-Hookean solids in detail to compare closely with previously published works. Nevertheless, we made the analyses for general incompressible hyperelastic solids. The analyses showed that the incompressible Mooney-Rivlin solids' crease characteristics are identical to those of incompressible neo-Hookean solids. The analyses with the Gent model revealed that the strain-stiffening in hyperelasticity enhances the far field's bracing capacity, altering the shape factor's dependence on the compressive strain to raise the crease resistance of the solid. The singular perturbation method, the crease-limit shape factor, and the TM-field energetics introduced in this work will help study the strength and the patterning of soft materials' surfaces and interfaces in various chemical or bio environments.

**CRediT authorship contribution statement**
Siyuan Song: Conceptualization, Methodology, Software, Data curation, Validation, Writing original draft. Mrityunjay Kothari: Early-stage conceptualization, Writing – review. Kyung-Suk Kim: Conceptualization, Methodology, Validation, Writing – review & editing, Supervision, Project administration.

**Declaration of Competing Interest**
The authors declare that they have no known competing financial interests or personal relationships that could have appeared to influence the work reported in this paper.

**Acknowledgments**
The authors gratefully acknowledge the supports provided by U.S. National Science Foundation (Grant CMMI-1934314) and U.S. Office of Naval Research (Grant N00014-22-1-2786). This article is dedicated to Professor Nick Triantafyllidis for his contributions to the stability theory of solids and structures on his 70th birthday.

**Figure Captions**

**Figure 1.** Incremental hyperelastic deformation: (a) A schematic of crease-field structures in a hyperelastic half-space under lateral compression. (b) Decomposition of affine finite deformation $(A) \rightarrow (B)$, Incremental deformation $(B) \rightarrow (C)$, and Tangential Manifold (TM) deformation $(C) \leftrightarrow (D)$; (E) Tangential stiffness projects the TM state in $(B) \leftrightarrow (D)$.

**Figure 2.** (a) An $M$ integral contour for a *bid* field in (B) frame. (b) configurational force balance diagram of a TM-ACG subsurface dislocation. (c) A diagram of a TM-ACG subsurface dislocation and its three image dislocations.

**Figure 3.** (a) Top insets: four types of singular-perturbation sources; Lower frame: an FEM analysis domain ($100 \times 100\ unit$) including a *bid* field and a range ($l = 37, t = 8\ unit$) of $M^{(TM)}$ domain integral. (b) $B_{(n)}$, $h_{(n)}$, and $S^{(bid)}_{(n)}$ variations in an iterative half-space analysis with finite-domain simulations for $1 - \lambda = 0.344$. (c) FEM-evaluated $M^{(TM)}_{\Gamma_\infty}(1 - \lambda)$ for loading and unloading. (d) *bid*-field shape factor, $S^{(bid)}(1 - \lambda)$, exhibiting creasing at $S^{(bid)}(0.356) = 1$.

**Figure 4.** FEM-simulated strain-energy density distributions near a surface flaw for $1 - \lambda =$ $(a)\ 0.1,\ (b)\ 0.2, (c)\ 0.3$, and (d) 0.344. Energy density distribution in the TM-ACG subsurface-dislocation field of $1 - \lambda = 0.344$ in (e) a neo-Hookean solid and (f) a Gent model of $J_m = 2$ for which crease does not occur up to the Gent limit.

**Figure 5.** (a) convergence of $\tilde{\sigma}^+_{11}(0, x_2) - \tilde{\sigma}^\infty_{11}$ to the leading-order field of $\sigma \sim 1/|x|^2$ (dashed line) at the far field; iterative FEM simulations (solid lines); TM-ACG subsurface-dislocation model analysis (a dash-dot line). (b) Incremental UPC-nominal stress distributions, $\tilde{\sigma}^-_{11}(0, x_2) - \tilde{\sigma}^\infty_{11}$, for $0 \leq |x_2|/B \leq 5$.

**Figure 6.** Dependence of (a) crease-limit shape factor $S^{(cr)}$ (b) crease-limit stretch $\lambda^{(cr)}$ on Gent's strain-stiffening parameter $J_m$.

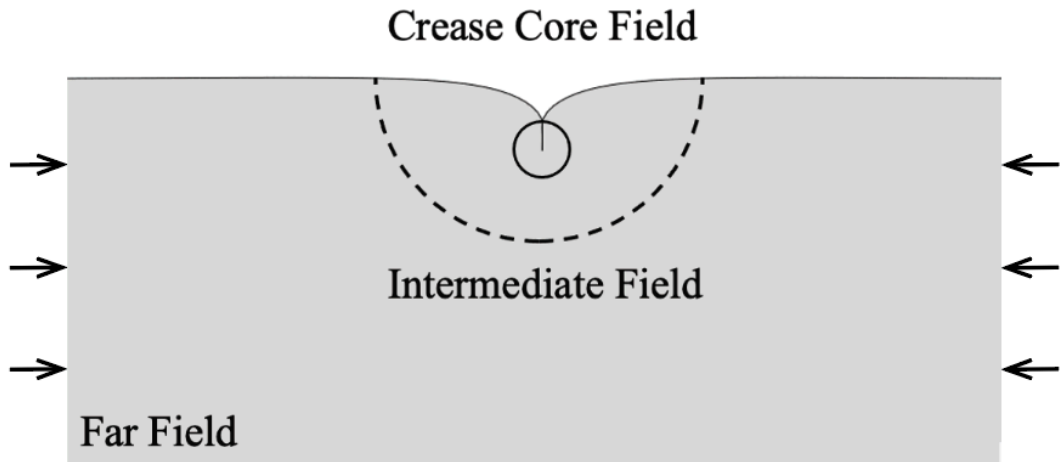

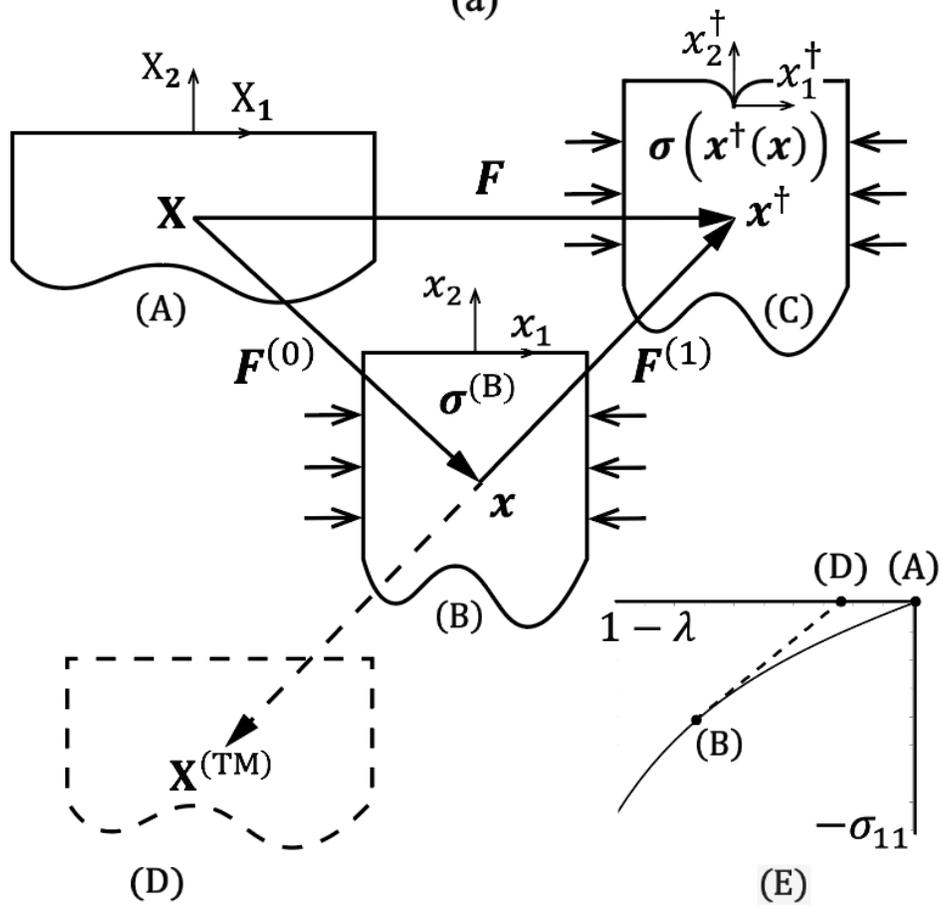

Figure 1.

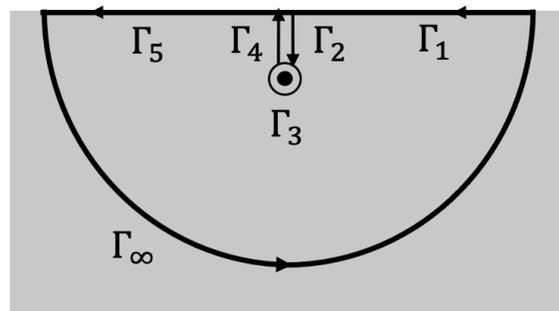

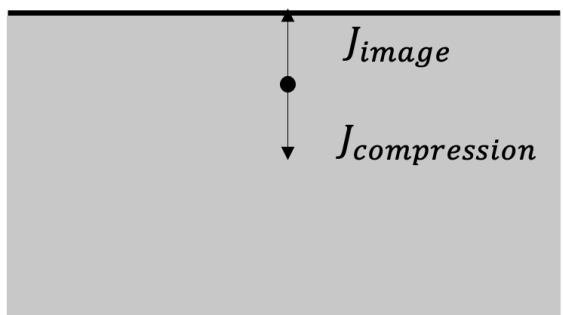

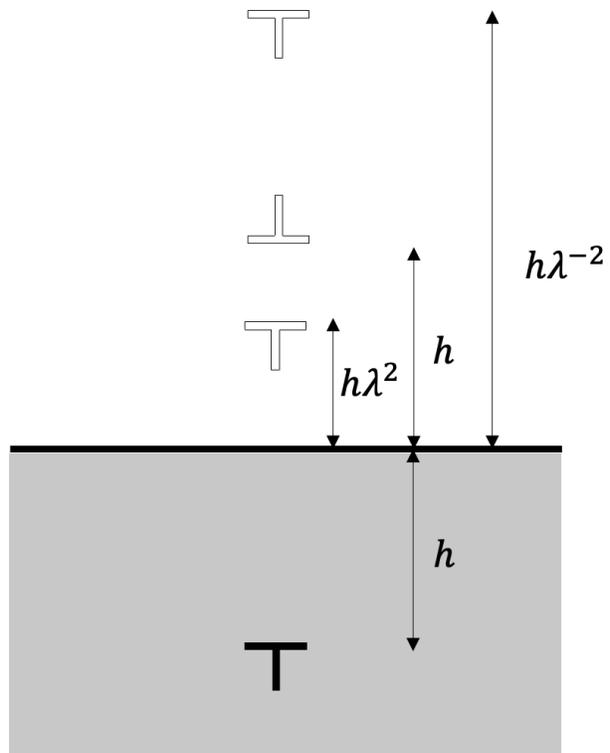

Figure 2.

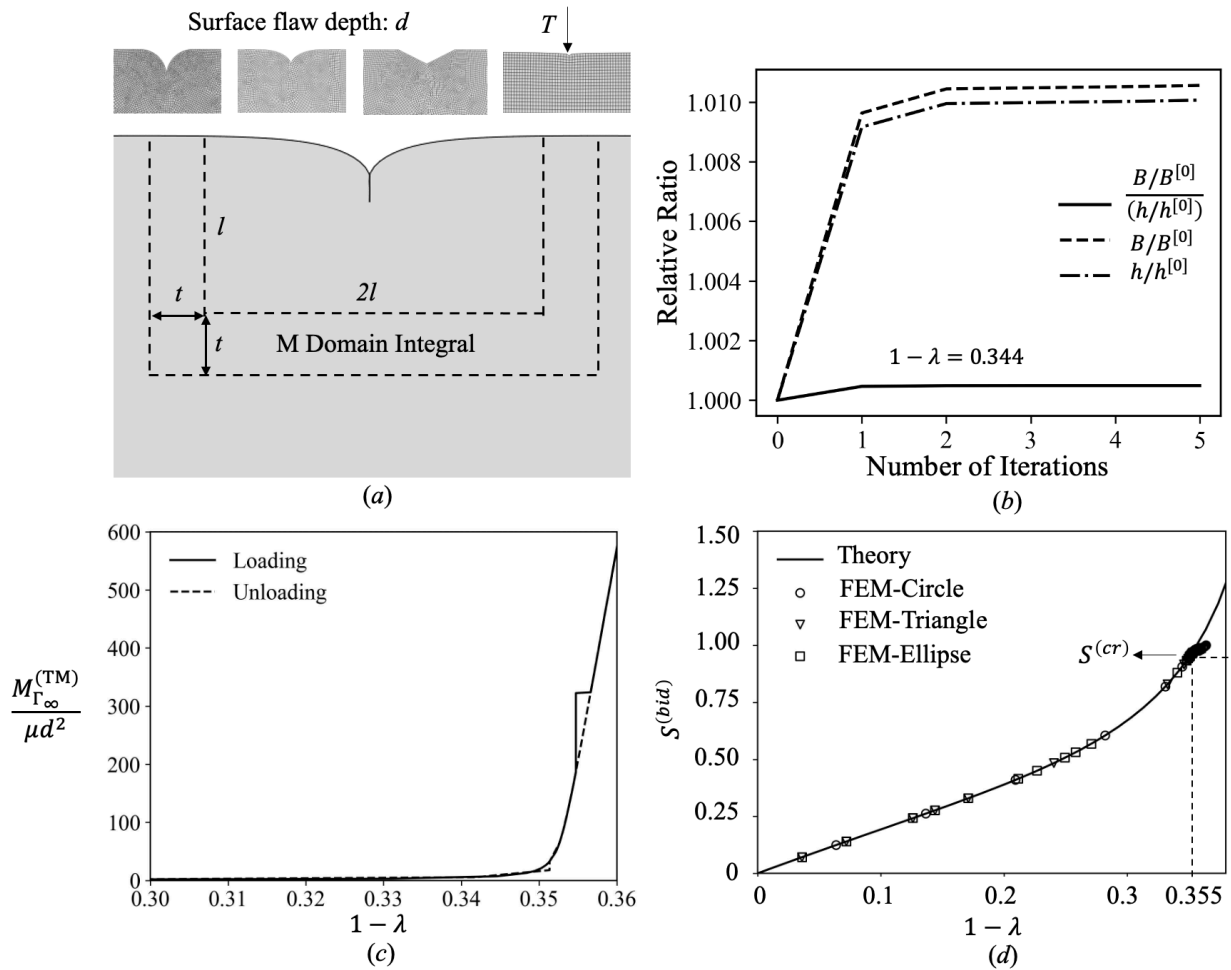

Figure 3.

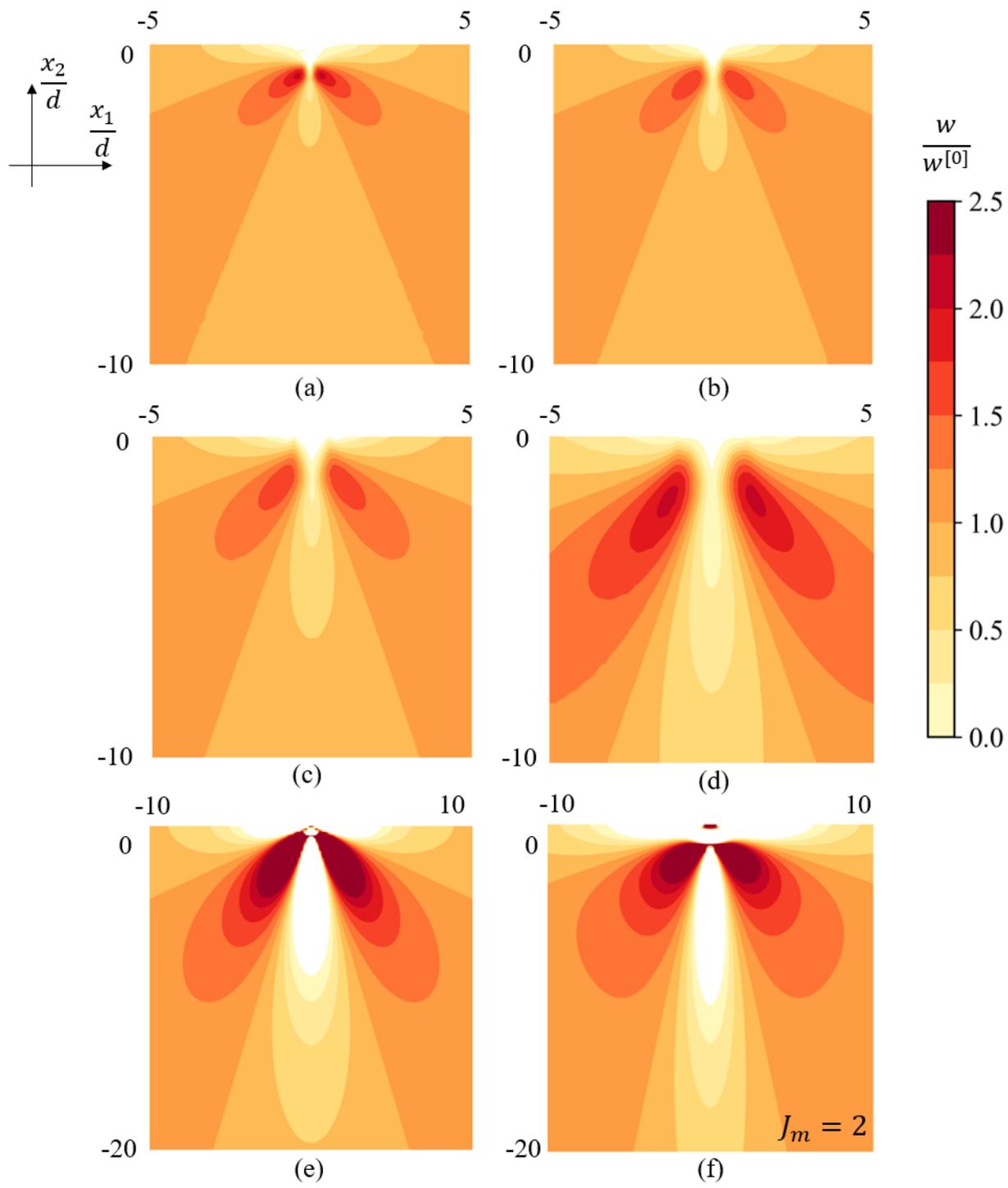

Figure 4.

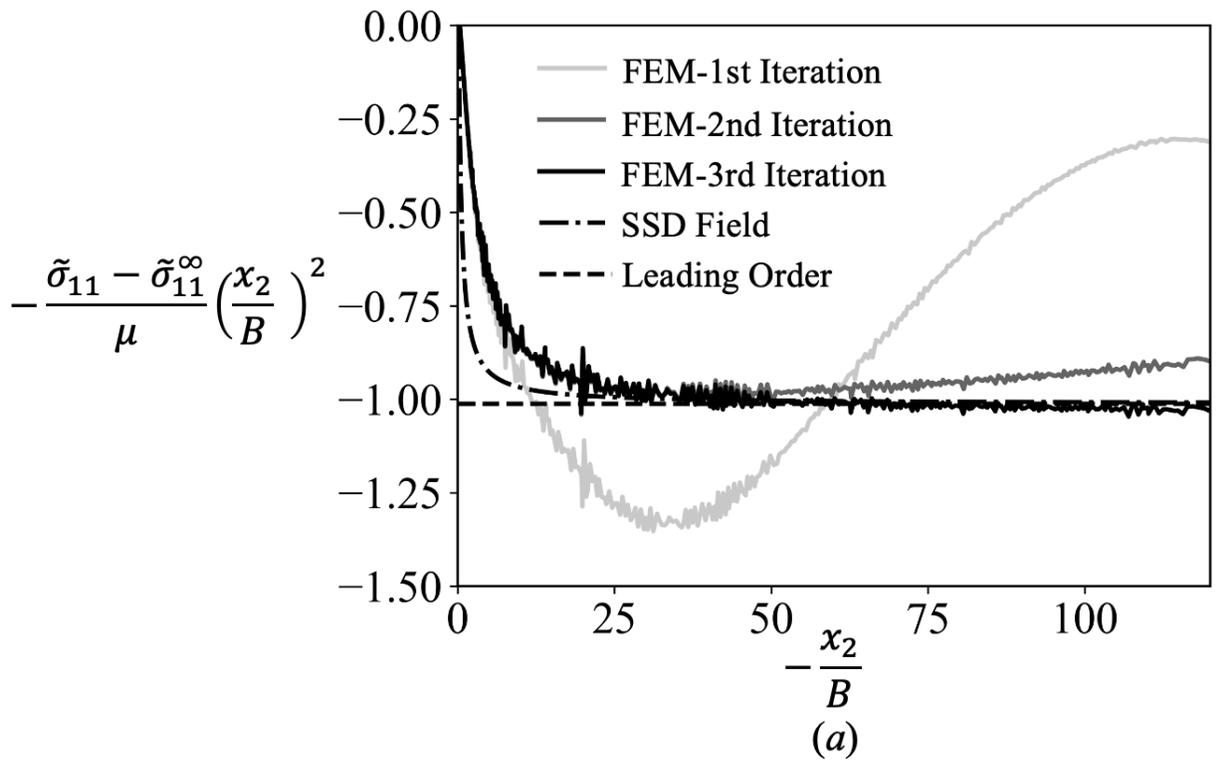
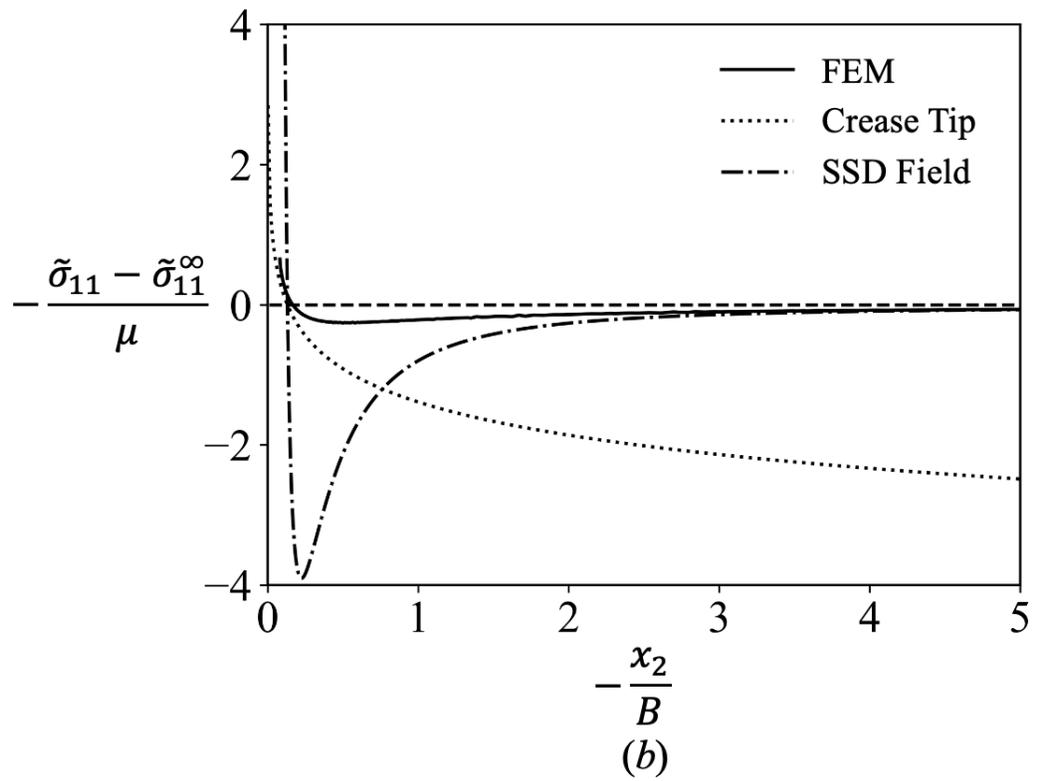

Figure 5.

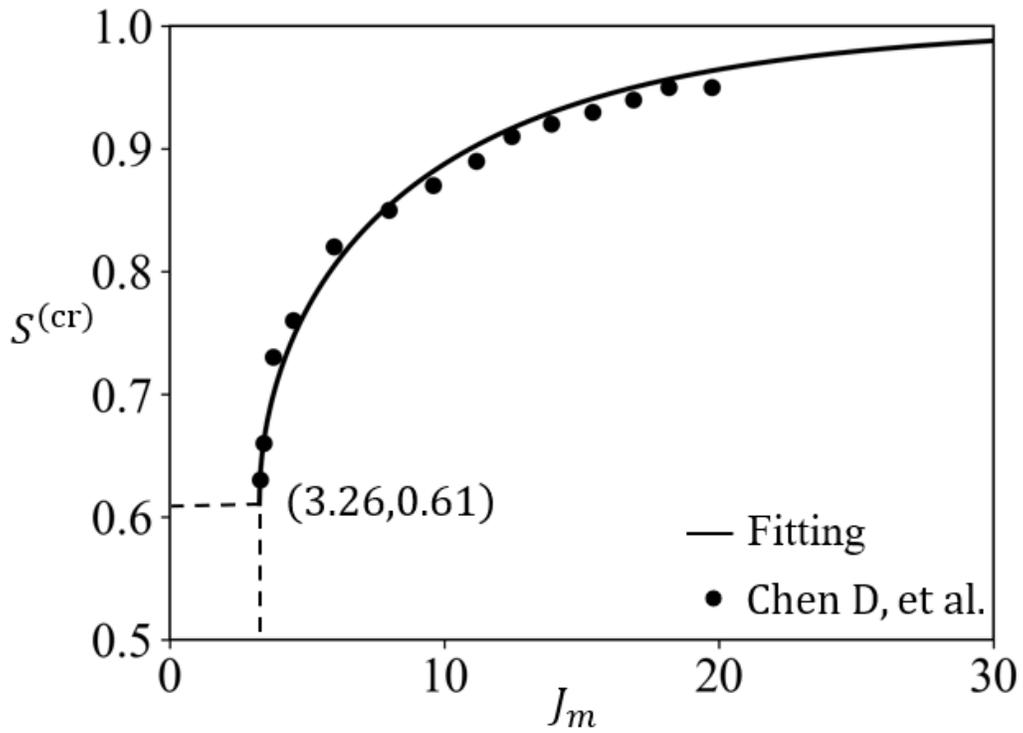

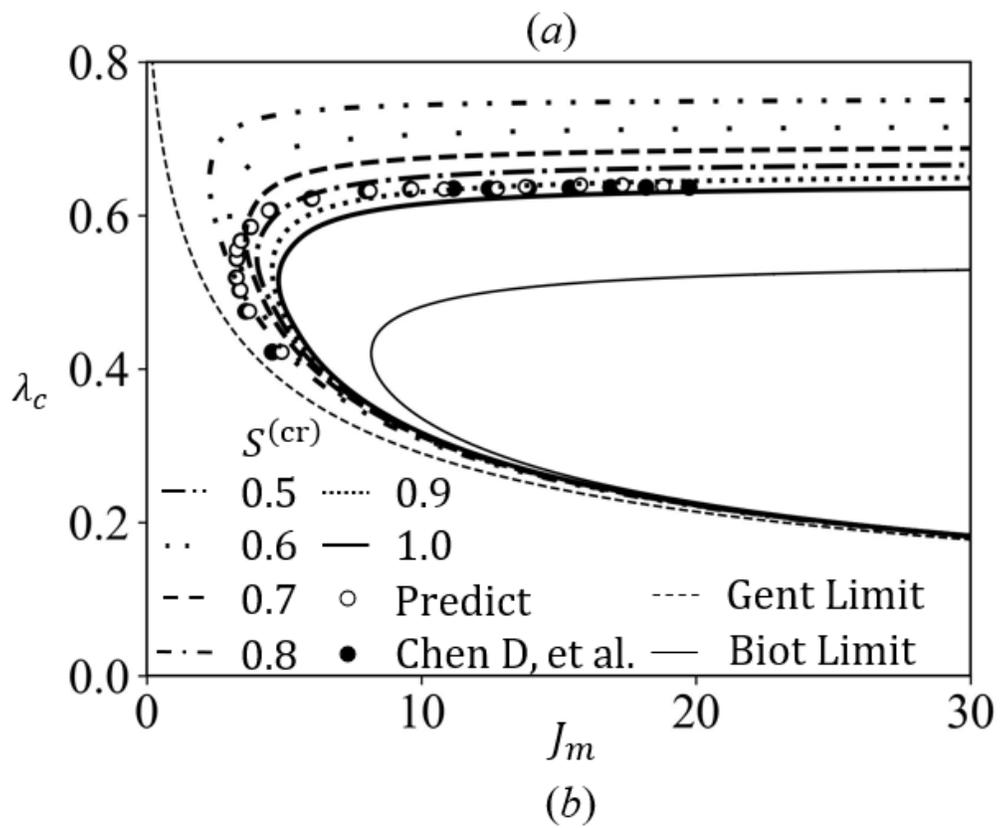

Figure 6

**Appendix A**: Path-Independence of $M_{\Gamma^{(B)}}^{(TM)}$ Integral

The tangential-compliance based TM *M*-integral is defined as

$$M_{\Gamma^{(B)}}^{(TM)} = \int_{\Gamma^{(B)}} x_\alpha \left[ n_\alpha \widehat{w}^{(c)}(\widetilde{\boldsymbol{\sigma}}) - n_\gamma \widetilde{\sigma}_{\beta\gamma} \hat{u}_{\beta,\alpha} \right] ds \tag{A1}$$

where the integration is evaluated in the $(x_1, x_2, x_3)$ frame, and $\Gamma^{(B)}$ is a closed path enclosing no singularities. According to the divergence theorem, we have

$$M_{\Gamma^{(B)}}^{(TM)} = \int_{\Gamma^{(B)}} \left[ \frac{\partial \widehat{w}^{(c)}}{\partial x_\alpha} x_\alpha + 2\widehat{w}^{(c)} - \left( \widetilde{\sigma}_{\beta\gamma} \frac{\partial \hat{u}_\beta}{\partial x_\gamma} + x_a \frac{\partial \widetilde{\sigma}_{\beta\gamma}}{\partial x_\gamma} \frac{\partial \hat{u}_\beta}{\partial x_\alpha} + x_a \widetilde{\sigma}_{\beta\gamma} \frac{\partial^2 \hat{u}_\beta}{\partial x_\alpha \partial x_\gamma} \right) \right] dA. \tag{A2}$$

Since $\frac{\partial \widehat{w}^{(c)}}{\partial x_\alpha} = \frac{\partial \widehat{w}^{(c)}}{\partial \hat{u}_{\beta,\gamma}} \frac{\partial \hat{u}_{\beta,\gamma}}{\partial x_\alpha} = \widetilde{\sigma}_{\beta\gamma} \frac{\partial^2 \hat{u}_\beta}{\partial x_\alpha \partial x_\gamma}$, $2\widehat{w}^{(c)} = \widetilde{\sigma}_{\beta\gamma} \frac{\partial \hat{u}_\beta}{\partial x_\gamma}$, and $\frac{\partial \widetilde{\sigma}_{\beta\gamma}}{\partial x_\gamma} = 0$, we have

$$M_{\Gamma^{(B)}}^{(TM)} = 0. \tag{A3}$$

**Appendix B**: Incremental Deformation in General Incompressible Hyperelastic Half-Space

For the general incompressible hyperelastic material, we can introduce a stream function $\psi$ for ($u_1 = \psi_{,2}$ and $u_2 = -\psi_{,1}$). According to the balance of force, $\psi$ satisfies the duel-coordinate $(v, \omega)$ harmonic,

$$\nabla_\omega^2 \nabla_v^2 \psi = 0 \tag{B1}$$

The two coordinates $v, w$ are represented as

$$v_1 = s_v x_1, \quad v_2 = x_2/s_v \tag{B2a}$$

$$\omega_1 = s_\omega x_1, \quad \omega_2 = x_2/s_\omega \tag{B2b}$$

where the coordinate-scaling factors $s_v, s_\omega$ can be determined by,

$$\begin{pmatrix} s_v \\ s_\omega \end{pmatrix} = \left( \frac{\lambda^2 k_1 + \lambda^{-2} k_3 \pm \sqrt{(\lambda^2 k_1 + \lambda^{-2} k_3)^2 - 4(k_2)^2}}{2\lambda^2 k_2} \right)^{\frac{1}{4}} \tag{B3}$$

with

$$k_1 = 4(\lambda^2 - \lambda^{-2}) \left[ \left( \frac{\partial^2 w}{\partial I_1 \partial I_1} + \frac{\partial^2 w}{\partial I_1 \partial I_2} \right) + (\lambda^{-2} + 1) \left( \frac{\partial^2 w}{\partial I_2 \partial I_1} + \frac{\partial^2 w}{\partial I_2 \partial I_2} \right) \right] + k_2 \tag{B4a}$$

$$k_2 = 2 \left( \frac{\partial w}{\partial I_1} + \frac{\partial w}{\partial I_2} \right) \tag{B4b}$$

$$k_3 = -4(\lambda^2 - \lambda^{-2}) \left[ \left( \frac{\partial^2 w}{\partial I_1 \partial I_1} + \frac{\partial^2 w}{\partial I_1 \partial I_2} \right) + (\lambda^2 + 1) \left( \frac{\partial^2 w}{\partial I_2 \partial I_1} + \frac{\partial^2 w}{\partial I_2 \partial I_2} \right) \right] + k_2. \tag{B4c}$$

The duel-coordinate harmonic functions can be articulated by analytic complex functions $f(\omega = \omega_1 + \omega_2 i)$ and $g(v = v_1 + v_2 i)$. Then, the incremental-displacement and the stress fields are expressed as

$$u_1 = \text{Re}\left( s_\omega^{-1} f(\omega) + s_v^{-1} g(v) \right) \tag{B5a}$$

$$u_2 = -\text{Im}\left( s_\omega f(\omega) + s_v g(v) \right) \tag{B5b}$$

$$\tilde{\sigma}_{11} = k_2 \left\{ (\lambda^2 - \lambda^{-2}) + \lambda^{-2} \text{Re} \left[ s_\omega^{-2} r_\omega \frac{\partial f}{\partial \omega} + s_v^{-2} r_v \frac{\partial g}{\partial v} \right] \right\} \tag{B6a}$$

$$\tilde{\sigma}_{12} = -\lambda^{-2} k_2 \text{Im} \left[ r_\omega \frac{\partial f}{\partial \omega} + r_v \frac{\partial g}{\partial v} \right] \tag{B6b}$$

$$\tilde{\sigma}_{21} = -\lambda^{-2} k_2 \text{Im} \left[ s_\omega^{-2} q_\omega \frac{\partial f}{\partial \omega} + s_v^{-2} q_v \frac{\partial g}{\partial v} \right] \tag{B6c}$$

$$\tilde{\sigma}_{22} = -\lambda^{-2} k_2 \operatorname{Re}\left[q_\omega \frac{\partial f}{\partial \omega} + q_v \frac{\partial g}{\partial v}\right]. \tag{B6d}$$

where

$$q_\alpha = 1 + \lambda^4 s_\alpha^4, \quad \alpha = \omega \text{ or } v \tag{B7a}$$

$$r_\alpha = s_\alpha^2 + s_\alpha^{-2}, \quad \alpha = \omega \text{ or } v. \tag{B7b}$$

The traction-free boundary conditions at $x_2 = 0$ require

$$\tilde{\sigma}_{12}(x_1, 0) = 0 \tag{B8a}$$

$$\tilde{\sigma}_{22}(x_1, 0) = 0. \tag{B8b}$$

Inserting (B7b&d) into (B8a&b), we see that $\frac{\partial f}{\partial \omega}(x_1, 0)$ and $\frac{\partial g}{\partial v}(x_1, 0)$ have nontrivial solutions if

$$q_\omega r_v - q_v r_\omega = 0 \tag{B9}$$

which determines the critical compressive strain for the surface-wrinkling bifurcation.

Furthermore, the TM-ACG subsurface dislocation solution is obtained as

$$f(\omega) = ia_2 \operatorname{Log}(\omega + h_1 i) + ia_4 \operatorname{Log}(\omega - h_2 i) + ia_6 \operatorname{Log}(\omega - h_3 i) \tag{B10a}$$

$$g(v) = ib_2 \operatorname{Log}(v + l_1 i) + ib_4 \operatorname{Log}(v - l_2 i) + ib_6 \operatorname{Log}(v - l_3 i) \tag{B10b}$$

where the coefficient can be determined by $h_2 = h_1 = h/s_\omega$, $l_2 = l_1 = h/s_v$, $h_3 = s_\omega h/s_v^2$, $l_3 = s_v h/s_\omega^2$, and

$$\frac{B}{2\pi} = \frac{a_2}{s_\omega} + \frac{b_2}{s_v} \tag{B11a}$$

$$\frac{a_2 q_\omega}{s_\omega} + \frac{b_2 q_v}{s_v} = 0 \tag{B11b}$$

$$a_4 = -\frac{Q^+}{Q^-} a_2 = \frac{s_\omega Q^+ R^-}{2 s_v Q^-} b_6 \tag{B11c}$$

$$a_6 = -\frac{2 s_\omega}{s_v Q^-} b_2 = \frac{2 s_\omega R^-}{s_v Q^- R^+} b_4 \tag{B11d}$$

with

$$Q^\pm = \frac{r_\omega}{r_v} \pm \frac{q_\omega}{q_v} \tag{B12a}$$

$$R^\pm = \frac{r_v}{r_\omega} \pm \frac{q_v}{q_\omega}. \tag{B12b}$$

The configurational force acting on the subsurface dislocation should be balanced and the condition for the configurational force equilibrium is derived as

$$k_2 B \left\{ (\lambda^2 - \lambda^{-2}) + \frac{B q_v q_\omega}{2\pi h \lambda^4 s^+ s^-} \left[ \frac{q_\omega}{\lambda^2 Q^- s_\omega^2} \left( \frac{Q^+}{2 q_\omega} - \frac{2 s_v^2}{q_v s^+} \right) - \frac{\lambda^{-2} q_v}{R^- s_v^2} \left( \frac{R^+}{2 q_v} - \frac{2 s_\omega^2}{q_\omega s^-} \right) \right] \right\} = 0 \tag{B13}$$

where

$$s^\pm = s_v^2 \pm s_\omega^2. \tag{B14}$$

The first term in $(B13)$ is the configurational force of the uniform compression field acting on the subsurface dislocation at $x_2 = -h$, the next two terms are of an image dislocation at $x_2 = h$, and the last two are of image dislocations at $x_2 = h s_\omega^2 / s_v^2$ and $h s_v^2 / s_\omega^2$, respectively.

The dislocation-core stiffness pre-factor in (23a) is derived as

$$H\left(\mathcal{S}_{\alpha\beta\zeta\eta}(\lambda)\right) = \frac{4\pi k_2}{\lambda^2 B} \left( a_4 \frac{r_\omega}{2 s_\omega} + b_4 \frac{r_v}{2 s_v} + a_6 \frac{r_\omega s_v^2}{s_\omega s^+} + b_6 \frac{r_v s_\omega^2}{s_v s^+} \right). \tag{B15}$$

**Appendix C**: Far-Field Incremental Deformation Caused by a Surface Flaw

in a Laterally-Compressed neo-Hookean Half-Space

In the limit of $\frac{h^2}{x_1^2+x_2^2\lambda^4} \ll 1$ (i.e., far field), the displacement and the stress fields are expressed as

$$u_1 = \frac{\Lambda^+}{\Lambda^-}\frac{(1-c^2\lambda^2)Bh}{\pi(c^2-1)}\left[\frac{x_1}{(x_1^2+\lambda^4 x_2^2)} - \frac{2x_1}{\Lambda^+(x_1^2+x_2^2)}\right] \qquad (C1a)$$

$$u_2 = \frac{\Lambda^+}{\Lambda^-}\frac{(1-c^2\lambda^2)Bh}{\pi(c^2-1)}\left[\frac{x_2}{(x_1^2+\lambda^4 x_2^2)} - \frac{2x_2}{\Lambda^+(x_1^2+x_2^2)}\right] \qquad (C1b)$$

$$\tilde{\sigma}_{11} = \mu\left(\lambda^2 - \frac{1}{\lambda^2}\right) - \frac{\Lambda^+}{\Lambda^-}\frac{(1-c^2\lambda^2)\mu Bh}{\pi(c^2-1)\lambda^2}\left[\frac{\Lambda^+(x_1^2-\lambda^4 x_2^2)}{(x_1^2+\lambda^4 x_2^2)^2} - \frac{4(x_1^2-x_2^2)}{\Lambda^+(x_1^2+x_2^2)^2}\right] \qquad (C2a)$$

$$\tilde{\sigma}_{22} = \frac{2\Lambda^+}{\Lambda^-}\frac{(1-c^2\lambda^2)\mu Bh}{\pi(c^2-1)\lambda^2}\left[\frac{x_1^2-\lambda^4 x_2^2}{(x_1^2+\lambda^4 x_2^2)^2} - \frac{x_1^2-x_2^2}{(x_1^2+x_2^2)^2}\right] \qquad (C2b)$$

$$\tilde{\sigma}_{12} = \frac{2\Lambda^+}{\Lambda^-}\frac{(1-c^2\lambda^2)\mu Bh}{\pi(c^2-1)\lambda^2}\left[-\frac{\Lambda^+ x_1 x_2}{(x_1^2+\lambda^4 x_2^2)^2} + \frac{4x_1 x_2}{\Lambda^+(x_1^2+x_2^2)^2}\right] \qquad (C2c)$$

$$\tilde{\sigma}_{21} = \frac{4\Lambda^+}{\Lambda^-}\frac{(1-c^2\lambda^2)\mu Bh}{\pi(c^2-1)}\left[-\frac{\lambda^2 x_1 x_2}{(x_1^2+\lambda^4 x_2^2)^2} + \frac{x_1 x_2}{\lambda^2(x_1^2+x_2^2)^2}\right] \qquad (C2d)$$

where

$$\Lambda^\pm = (1 \pm \lambda^4). \qquad (C3)$$